\documentstyle[12pt, epsf]{article}

\advance \topmargin by -\headheight
\advance \topmargin by -\headsep

\evensidemargin 
\oddsidemargin

 \def\ex{{\hbox{\rm e}}}

\def\im{{\hbox{\rm Im}}}  \def\tr{{\hbox{\rm 
Tr}}}

\def\pff{{\hbox{\rm Pf}}}

\def\ie{{\em i.e.}}
\def\be{\begin{equation}}
\def\ee{\end{equation}}
\def\bea{\begin{eqnarray}}
\def\eea{\end{eqnarray}}

\def\im{{\hbox{\rm Im}}}
\def\ker{{\hbox{\rm ker}}}

\def\cale{\cal{E}}
\def\ad{\hbox{\rm ad}}

\def\intl{\int\limits}
\def\ex{{\hbox{\rm e}}}  
\def\tr{{\hbox{\rm Tr}}}
\def\too{\longrightarrow}
\def\half{{1\over 2}} 

\def\to{\rightarrow}

\def\sqr#1#2{{\vcenter{\vbox{\hrule height.#2pt
  \hbox{\vrule width.#2pt height#1pt \kern#1pt
    \vrule width.#2pt}
  \hrule height.#2pt}}}}

\def\deriv{{\cal D}}
\def\dalpha{{\dot\alpha}}

\newcommand{\CS}{{\scriptstyle {\rm CS}}}
\newcommand{\CSs}{{\scriptscriptstyle {\rm CS}}}
\newcommand{\ZZ}{{\mbox{{\bf Z}}}}
\newcommand{\RR}{{\mbox{{\bf R}}}}

\newcommand{\topo}{{\scriptstyle {\rm top}}}

\begin{document}
\begin{titlepage}

\begin{flushright} CERN-TH/97-250 \\ US-FT-30/97 \\ hep-th/9709192 \\
September 1997
\end{flushright}

\vspace*{20pt}
\bigskip
\begin{center}
 {\Large \bf Lectures on Topological Quantum Field Theory\footnotemark}
\vskip 0.9truecm

\footnotetext{Lectures given at the workshop ``Trends in Theoretical
Physics", CERN-Santiago de Compostela-La Plata Meeting,  held in La Plata,
Argentina, April 28-May 6 1997.}

{J. M. F. Labastida$^{a,b}$ and  Carlos Lozano$^{b}$}

\vspace{1pc}

{\em $^a$ Theory Division, CERN,\\
 CH-1211 Geneva 23, Switzerland.\\
 \bigskip
  $^b$ Departamento de F\'\i sica de Part\'\i culas,\\ Universidade de
Santiago de Compostela,\\ E-15706 Santiago de Compostela, Spain.\\}

\vspace{5pc}

{\large  Abstract}
\end{center}

\vspace{2pc}

In these lectures we present a general introduction to topological  quantum
field theories. These theories are discussed in the framework  of the
Mathai-Quillen formalism and in the context of twisted $N=2$ supersymmetric
theories. We discuss in detail the recent developments  in Donaldson-Witten
theory obtained from the application of results based on duality for $N=2$
supersymmetric Yang-Mills theories. This involves  a description of the
computation of Donaldson invariants in terms of  Seiberg-Witten invariants.
Generalizations of Donaldson-Witten theory are reviewed, and the
structure of the vacuum expectation values of their observables  is
analyzed in the context of duality for the simplest case.

\normalsize\baselineskip=15pt
\setcounter{footnote}{0}
\renewcommand{\thefootnote}{\alph{footnote}}

\end{titlepage}

\def\theequation{\thesection.\arabic{equation}}

\tableofcontents

\vspace{1pc}


\section{Introduction}
\setcounter{equation}{0}

Topological quantum field theory (TQFT) emerged in the eighties as a new 
relation between mathematics and physics. This relation connected some of
the most advanced ideas in the two fields. The nineties have been
characterized by its development, originating unexpected results in
topology and testing some of the most fundamental ideas in quantum field
theory and string theory.

The first TQFT was formulated by Witten \cite{tqft} in 1988. He constructed
the theory now known as Donaldson-Witten theory, which constitutes a
quantum field theory representation of the theory of Donaldson
invariants \cite{donaldcero,donald}. His work was strongly influenced by M.
Atiyah \cite{atiyah}. In 1988 Witten formulated also another two TQFTs
which have been widely studied during the last ten years: topological
sigma models in two dimensions \cite{tsm} and Chern-Simons gauge
theory \cite{csgt} in three dimensions. These theories are related,
respectively, to Gromov invariants \cite{gromov}, and to knot and link
invariants as the Jones polynomial \cite{jones} and its generalizations. 

TQFT has provided an entirely new approach to study topological invariants.
Being a quantum field theory, TQFT can be analyzed from different points of
view. The richness inherent to quantum field theory can be exploited to
obtain different perspectives on the topological invariants involved in
TQFT. This line of thought has shown to be very successful in the last
years and new topological invariants as well as new relations between them
have been obtained.

TQFTs have been studied from both, perturbative and non-perturbative
points of view. In the case of Chern-Simons gauge theory, non-perturbative
methods have been applied to obtain properties \cite{csgt,gambini} of knot
and link invariants, as well as general procedures for their
computation \cite{martin,kaul,torus}. Perturbative methods have also been
studied for this theory \cite{gmm,corrida,natan,alla,esther} providing integral
representations for Vassiliev \cite{vass}
invariants \cite{kont,barnatan,torusknots,factori}. In Donaldson-Witten
theory perturbative methods have proved its relation to Donaldson
invariants. Non-perturbative methods have been applied \cite{abm} after the
work by Seiberg and Witten \cite{sw} on $N=2$ supersymmetric Yang-Mills
theory. The outcome of this application is a totally unexpected relation
between Donaldson invariants and a new set of topological invariants
called Seiberg-Witten invariants. One of the main purposes of these
lectures is to describe the general aspects of this result.

Donaldson-Witten theory is a TQFT of cohomological type. TQFTs of this
type can be formulated in a variety of frameworks. The most geometric one
corresponds to the Mathai-Quillen formalism \cite{mathai}. In this
formalism a TQFT is constructed out of a moduli problem \cite{jeffrey}.
Topological invariants are then defined as integrals of a certain Euler
class (or wedge products of the Euler class with other forms) over the
resulting moduli space. A different framework is the one based on the
twisting of $N=2$ supersymmetry. In this case, information on the physical
theory can be used in the TQFT. Indeed, it has been in this framework
where Seiberg-Witten invariants have shown up.  After Seiberg and Witten
worked out the low energy effective action of $N=2$ supersymmetric
Yang-Mills theory it became clear that a twisted version of this effective
action could lead to topological invariants related to Donaldson
invariants. These twisted actions \cite{rocek,top,ans} revealed a new
moduli space, the moduli space of abelian monopoles \cite{abm}. Its
geometric structure has been derived in the context of the Mathai-Quillen
formalism \cite{abmono}. Invariants associated to this moduli space should
be related to Donaldson invariants. This turned out to be the case. The
relevant invariants for the case of $SU(2)$ as gauge group are the
Seiberg-Witten invariants.

Donaldson-Witten theory has been generalized  after studying its coupling
to topological matter fields \cite{rocek,top,ans}. The resulting theory can
be regarded as a twisted form of $N=2$ supersymmetric Yang-Mills theory
coupled to hypermultiplets, or, in the context of the Mathai-Quillen
formalism, as the TQFT associated to the moduli space of non-abelian
monopoles \cite{nabm}. Perturbative and non-perturbative methods have been
applied to this theory for the case of
$SU(2)$ as gauge group and one hypermultiplet of matter in the fundamental
representation \cite{last}. In this case, again, it turns out that the
generalized Donaldson invariants can be written in terms of Seiberg-Witten
invariants. It is not known at the moment which one is the situation for
other groups and representations but one would expect that in general the
invariants associated to non-abelian monopoles could be expressed in terms
of some other simpler invariants, being Seiberg-Witten invariants just the
first subset of the full set of invariants.

In Table \ref{latabla} we have depicted the present situation in three and
four dimensions relative to Chern-Simons gauge theory and Donaldson-Witten
theory, respectively. These theories share some common features. Their
topological invariants are labeled with group-theoretical data: Wilson
lines for different representations and gauge groups (Jones polynomial and
its generalizations), and non-abelian monopoles for different
representations and gauge groups (generalized Donaldson polynomials);
these invariants can be written in terms of topological invariants which
are independent of the group and representation chosen: Vassiliev
invariants and Seiberg-Witten invariants. This structure leads to the idea
of universality classes \cite{last,tesis} of topological invariants. In
this respect Vassiliev invariants constitute a class in the sense that all
Chern-Simons or quantum group knot invariants for semi-simple groups can
be expressed in terms of them. Similarly, Seiberg-Witten invariants
constitute another class since generalized Donaldson invariants associated
to several moduli spaces can be written in terms of them. This certainly
holds for the two cases described above but presumably it holds for other
groups. It is very likely that Seiberg-Witten invariants are the first set
of a series of invariants, each defining a universality class.

\begin{table}[htbp]
\centering
\begin{tabular}{|c|c|c|}  \hline
 & $d=3$  &  $d=4$ \\  \hline perturbative & Vassiliev & Donaldson \\
\hline non-perturbative & Jones & Seiberg-Witten \\ \hline
\end{tabular}
\caption{Topological invariants in the perturbative and the
non-perturbative regimes for $d=3$ and $d=4$.}
\label{latabla}
\end{table}

These lectures are organized as follows. In sect. 2 we present a general
introduction to TQFT from a functional integral point of view. In sect. 3
we review the Mathai-Quillen formalism and we discuss it in the context of
supersymmetric quantum mechanics and topological sigma models. In sect. 4
we introduce Donaldson-Witten theory in the framework of the Mathai-Quillen
formalism, and from the point of view of twisting $N=2$ supersymmetric
Yang-Mills theory. We then discuss the computation of its observables from
a perturbative point of view, showing its relation to Donaldson
invariants, and then from a non-perturbative one obtaining its expression
in terms of Seiberg-Witten invariants. This last analysis is done in two
different approaches: in the abstract approach, based on the structure of
$N=1$ supersymmetric Yang-Mills theory and valid only for K\"ahler
four-manifolds with
$H^{(2,0)}\ne 0$, and in the concrete approach, based on the structure of
$N=2$ supersymmetric Yang-Mills theory and valid for any four-manifold.
Explicit expressions for Donaldson invariants are collected for the case of
$SU(2)$ as gauge group and simply-connected four-manifolds with $b_2^+>1$.
In sect. 5 we describe the generalizations of Donaldson-Witten theory and 
review, for simply-connected four-manifolds with $b_2^+>1$, the structure
of the vacuum expectation values of its observables for the case of
$SU(2)$ as gauge group and one hypermultiplet in the fundamental
representation. Finally, in sect. 6 we include some final remarks.

Before entering into the core of these lectures let us recall that
excellent reviews on TQFTs are already
available \cite{thompson,moore,blau,freed}. In these lectures we have
mainly concentrated on subjects not covered in those reviews though,
trying to be self-contained, some overlapping is unavoidable. There are
also good reviews on Seiberg-Witten
invariants \cite{donaldreview,matilde,morgan,yang} from a mathematical
perspective.

\vfill
\newpage

\section{Topological Quantum Field Theory}
\setcounter{equation}{0}

In this section we present the general structure of TQFT from a  functional
integral point of view. As in ordinary quantum field theory, the 
functional integration involved is not in general well defined. Similarly
to that  case  this has led to the construction of an  axiomatic approach
\cite{axio}. In these lectures, however, we are not going to describe this
approach. We will concentrate on the functional integral point of view.
Although not well defined in general, this is the  approach which has
shown to be more successful.

Our basic topological space will be an $n$-dimensional Riemannian 
manifold $X$ endowed with a metric $g_{\mu\nu}$. On this space we will
consider  a  set of fields $\{\phi_i\}$, and  a real functional of these
fields, $S(\phi_i)$, which will be regarded as the action of the  theory.
We will  consider operators,
${\cal O}_\alpha(\phi_i)$, which will be in general arbitrary  functionals
of the fields. In TQFT these functionals are labeled by  some set of
indices
$\alpha$ carrying topological or group-theoretical data. The vacuum
expectation value (vev) of a product  of these operators is defined as the
following functional integral: 
\begin{equation}
\langle {\cal O}_{\alpha_1} {\cal O}_{\alpha_2} \cdots {\cal O}_{\alpha_p}
\rangle =
\int [D\phi_i]  {\cal O}_{\alpha_1}(\phi_i) {\cal  O}_{\alpha_2}(\phi_i)
 \cdots {\cal O}_{\alpha_p}(\phi_i) \exp\big(-S(\phi_i)\big).
\label{sanse}
\end{equation} A quantum field theory is considered topological if it
possesses the following property:
\begin{equation}
\frac{\delta}{\delta g^{\mu\nu}} 
\langle {\cal O}_{\alpha_1} {\cal O}_{\alpha_2} \cdots {\cal O}_{\alpha_p}
\rangle = 0,
\label{reme}
\end{equation}
\ie, if the vacuum expectation values (vevs) of some set of selected  
operators remain invariant under variations of the metric $g_{\mu\nu}$  on
$X$. If such is the case these operators are called observables.

There are two ways to guarantee, at least formally, that condition  
(\ref{reme}) is satisfied. The first one corresponds to the situation in
which both,  the action, S, as well as the operators ${\cal O}_{\alpha}$,
are metric independent. These TQFTs are called of {Schwarz}
type \cite{thompson}. In the case of Schwarz-type theories one must first 
construct an action which is independent of the metric $g_{\mu\nu}$. The
method is best  illustrated by considering an example. Let us take into
consideration the most  interesting case of this type of theories:
Chern-Simons gauge theory \cite{csgt}.  The data in Chern-Simons gauge
theory are the following: a  differentiable compact three-manifold $M$, a
gauge group
$G$, which will be taken  simple and compact, and an integer parameter
$k$. The action is the integral of the Chern-Simons form associated to a 
gauge connection $A$ corresponding to the group $G$:
\begin{equation} S_{\CS} (A) = \int_M \tr (A\wedge d A + \frac{2}{3}
A\wedge A\wedge A).
\label{valery}
\end{equation}

Observables are constructed out of operators which do not contain the 
metric
$g_{\mu\nu}$. In gauge invariant theories, as it is the case, one must 
also demand for these operators invariance under gauge transformations. An
important set of observables in Chern-Simons gauge theory is  constituted
by the trace of the holonomy of the gauge connection $A$ in  some
representation
$R$ along a 1-cycle $\gamma$, \ie, the Wilson  loop:
\begin{equation}
\tr_R \big( {\hbox{\rm Hol}}_\gamma (A) \big) =
\tr_R {\hbox{\rm P}} \exp \int_\gamma A.
\label{silvie}
\end{equation} The vevs are labeled by representations, $R_i$, and  
embeddings, $\gamma_i$, of $S^1$ into $M$:
\begin{equation}
\langle \tr_{R_1} {\hbox{\rm P}} \ex^{\int_{\gamma_1} A}
  \dots
  \tr_{R_n} {\hbox{\rm P}} \ex^{\int_{\gamma_n} A} \rangle 
  \nonumber \\ = \int [DA] \tr_{R_1} {\hbox{\rm P}}
\ex^{\int_{\gamma_1} A}
  \dots
  \tr_{R_n} {\hbox{\rm P}} \ex^{\int_{\gamma_n} A}
  \ex^{\frac{i k}{4\pi} S_{\CSs} (A) }.
\label{encarna}
\end{equation} A non-perturbative analysis of Chern-Simons gauge
theory \cite{csgt}  shows that the invariants associated to the observables
(\ref{encarna})  are  knot and link invariants of polynomial type as  the
Jones polynomial \cite{jones} and its generalizations (HOMFLY \cite{homfly},
Kauffman \cite{kauffman},  Akutsu-Wadati \cite{aku}, etc.). The perturbative
analysis \cite{gmm,natan,alla,esther} has also led to this result and has  
shown to provide a very useful framework to study Vassiliev  
invariants \cite{kont,bilin,barnatan,torusknots,factori} (see
 \cite{preport} for a brief review).

An important set of theories of Schwarz type are the BF
theories \cite{horo}. These theories can be formulated in any dimension and
are considered, as Chern-Simons gauge theory, exactly solvable quantum
field theories. We will not describe them in these lectures. They have
acquired importance recently since it has been pointed out that
four-dimensional Yang-Mills theories could be regarded as a deformation of
these theories \cite{cottados}. It is important also to remark that
Chern-Simons gauge theory plays an important role in the context of
the Ashtekar approach \cite{ashtekar} to the quantization of
four-dimensional gravity \cite{baez,pullin}.

The second way to guarantee (\ref{reme}) corresponds to the case in  which
there exists a symmetry, whose infinitesimal form will be denoted by
$\delta$, satisfying the  following properties:
\begin{equation}
\delta {\cal O}_{\alpha}(\phi_i) = 0, 
\;\;\;\;\; T_{\mu\nu}(\phi_i) = \delta G_{\mu\nu}(\phi_i),
\label{angela}
\end{equation} where $T_{\mu\nu}(\phi_i)$ is the energy-momentum tensor of
the theory,
\ie,
\begin{equation} T_{\mu\nu} (\phi_i) = \frac{\delta}{\delta g^{\mu\nu}}
S(\phi_i),
\label{rosi}
\end{equation} and $G_{\mu\nu}(\phi_i)$ is some tensor.

The fact that $\delta$ in (\ref{angela}) is a symmetry of the theory 
means that the transformations, $\delta\phi_i$, of the fields are such
that  both,
$\delta S(\phi_i) =0$, and, $\delta {\cal O}_{\alpha} (\phi_i) =0 $.  
Conditions (\ref{angela}) lead, at least formally, to the following 
relation for vevs:
\begin{eqnarray} & &\frac{\delta}{\delta g^{\mu\nu}} 
\langle {\cal O}_{\alpha_1} {\cal O}_{\alpha_2} \cdots {\cal O}_{\alpha_p}
\rangle  = -\int [D\phi_i]  {\cal O}_{\alpha_1}(\phi_i) {\cal
O}_{\alpha_2}(\phi_i)
 \cdots {\cal O}_{\alpha_p}(\phi_i) T_{\mu\nu}  
\exp\big(-S(\phi_i)\big)
\nonumber \\ & & \,\,\,\,\,\,\, \,\,\,\,\,\,\, = -\int [D\phi_i]
\delta\Big( {\cal O}_{\alpha_1}(\phi_i) {\cal O}_{\alpha_2}(\phi_i)
 \cdots {\cal O}_{\alpha_p}(\phi_i) G_{\mu\nu}  
\exp\big(-S(\phi_i)\big)\Big)
 =  0,
\label{rosina}
\end{eqnarray} which implies that the quantum field theory can be regarded
as topological. In (\ref{rosina}) it has been assumed that the action and
the measure $[D\phi_i]$ are invariant under the symmetry  
$\delta$. We have assumed also in (\ref{rosina}) that the observables are
metric-independent. This is a common situation in this type of  theories,
but it does not have to be necessarily so. In fact, in view of
(\ref{rosina}),  it would be possible to consider a wider class of
operators satisfying:
\be {\delta\over{\delta g_{\mu\nu}}}{\cal O}_{\alpha}(\phi_i)=
\delta O^{\mu\nu}_{\alpha}(\phi_i),
\label{queso}
\ee with $O^{\mu\nu}_{\alpha}(\phi_i)$ a certain functional of the fields 
of the theory.

This second type of TQFTs are called cohomological or of {Witten}
type \cite{thompson,coho}. One of its main representatives is
Donaldson-Witten theory \cite{tqft}, which can be regarded as a certain
{\it twisted} version of
$N=2$ supersymmetric Yang-Mills theory. It is important to remark that the
symmetry $\delta$ must be a scalar symmetry. The reason is that, being a
global  symmetry, the corresponding  parameter must be covariantly
constant and for arbitrary manifolds this  property, if it is satisfied at
all, implies strong restrictions unless  the parameter is a scalar.

Most of the TQFTs of cohomological type satisfy the relation:
\be S(\phi_i)=\delta\Lambda(\phi_i),
\label{caca}
\ee for some functional $\Lambda(\phi_i)$. This has far-reaching
consequences, for it means that the topological  observables of the theory
(in particular the partition function itself) are  independent of the
value of the coupling constant. Indeed, let us consider for example  the
vev: 
\begin{equation}
\langle {\cal O}_{\alpha_1} {\cal O}_{\alpha_2} \cdots {\cal O}_{\alpha_p}
\rangle =
\int [D\phi_i]  {\cal O}_{\alpha_1}(\phi_i) {\cal  O}_{\alpha_2}(\phi_i)
 \cdots {\cal O}_{\alpha_p}(\phi_i)  
\exp\big(-{1\over{g^2}}S(\phi_i)\big).
\label{sansed}
\end{equation} Under a change in the coupling constant, $1/g^2\to
1/g^2-\Delta$, one  has (assuming that the observables do not depend on
the coupling), up to first order in $\Delta$:
\bea
\langle {\cal O}_{\alpha_1} {\cal O}_{\alpha_2} \cdots {\cal O}_{\alpha_p}
\rangle&\too& 
\langle {\cal O}_{\alpha_1} {\cal O}_{\alpha_2} \cdots {\cal O}_{\alpha_p}
\rangle  \nonumber\\ &+&\Delta
\int [D\phi_i] \delta\left[ {\cal O}_{\alpha_1}(\phi_i) {\cal
O}_{\alpha_2}(\phi_i)
 \cdots {\cal O}_{\alpha_p}(\phi_i) 
\Lambda(\phi_i)\exp\big(-{1\over{g^2}}S(\phi_i)\big)\right]\nonumber\\&=  &
\langle {\cal O}_{\alpha_1} {\cal O}_{\alpha_2} \cdots {\cal O}_{\alpha_p}
\rangle.
\label{sunset}
\eea Hence, observables can be computed either in the weak coupling limit, 
$g\to 0$, or in the strong coupling limit, $g\to\infty$. 

So far we have presented a rather general definition of TQFT and made a
series of  elementary remarks. Now we will analyze some aspects of its
structure. We begin pointing out that given a theory in which
(\ref{angela}) holds one can build correlators which correspond to 
topological invariants (in the sense that they are invariant under
deformations  of the metric  
$g_{\mu\nu}$) just by considering the operators of the theory which are
invariant  under the symmetry. We will call these operators observables.
Actually, to be  more precise, we will call observables to certain classes
of those  operators. In virtue of eq. (\ref{rosina}),  if one of these 
operators can be  written as a symmetry transformation of another
operator, its presence in a  correlation function will make it vanish.
Thus we may identify operators  satisfying (\ref{angela}) which differ by
an operator which corresponds to a  symmetry transformation of another
operator. Let us denote the set of the resulting classes by $\{\Phi\}$. Actually, in
general, one could identify  bigger sets of operators since two operators
of which one of them does not  satisfy (\ref{angela}) could lead to the
same invariant if they differ by an  operator which is a symmetry
transformation of another operator. For example,  consider
${\cal O}$ such that $\delta{\cal O}=0$ and
${\cal O}+\delta\Gamma$. Certainly, both operators lead to the same
observables. But it may well happen that $\delta^2\Gamma\neq 0$ and  therefore
we have operators which do not satisfy (\ref{angela}) that must be 
identified. The natural way out is to work {\it equivariantly}, which in
this  context means that one must consider only operators which are
invariant under both,  
$\delta$ and $\delta^2$. In turns out that in most of the cases (and in  
particular, in all  the cases that we will be considering) $\delta^2$ is a
gauge  transformation,  so in the end all that has to be done is to
restrict the analysis to gauge-invariant operators, a very natural
requirement. Hence, by  restricting the analysis to the  appropriate set
of operators, one has that in  fact,  
\be 
\delta^2=0. 
\label{once}
\ee

Property (\ref{once}) has striking consequences on the features of TQFT.
First, the symmetry must be odd which implies the presence in the theory
of commuting and anticommuting fields. For example, the tensor
$G_{\mu\nu}$ in (\ref{angela}) must be  anticommuting. This is the first
appearance of an odd non-spinorial field in TQFT. Those kinds of objects
are standard features of cohomological TQFTs.  Second, if  we denote by
$Q$ the operator which implements this symmetry, the  observables of the
theory can be described as the cohomology classes of $Q$: 
\be
\{\Phi\} = {\ker \, Q\over \im\, Q},\,\,\,\,\,\,\,\,\,\,\,\,\,\,\,  Q^2=0.
\label{doce}
\ee

Equation (\ref{angela}) means that in addition to the Poincare group  the
theory possesses a symmetry generated by an odd version of the Poincare
group. The corresponding odd generators are constructed out of the tensor
$G_{\mu\nu}$ in much the same way as the ordinary Poincare generators are
built out of $T_{\mu\nu}$. For example, if $P_\mu$ represents the ordinary
momentum operator, there exists a corresponding odd one $G_\mu$ such that,
\be P_\mu = \{Q, G_\mu\}.
\label{trece}
\ee

Let us discuss the structure of the Hilbert space of the theory in virtue
of the symmetries that we have just described. The states of  this space
must correspond to representations of the algebra generated by  the
operators in the Poincare groups and by $Q$. Furthermore, as follows
from our analysis of operators leading to (\ref{doce}), if one is 
interested only in states
$|\Psi\rangle$ leading to topological  invariants one must consider states
which satisfy, 
\be Q|\Psi\rangle=0,
\label{catorce}
\ee and two states which differ by a $Q$-exact state must be identified.
The odd Poincare group can be used to generate descendant states out  of a
state satisfying (\ref{catorce}).  The operators $G_\mu$  act
non-trivially on the states and in fact, out of a state satisfying
(\ref{catorce}) we can build additional states using this generator.  The
simplest case consists of,  
\be
\int_{\gamma_1} G_{\mu} |\Psi\rangle, 
\label{quince}
\ee  where $\gamma_1$ is a 1-cycle. One can easily verify using  
(\ref{angela}) that this new state satisfies (\ref{catorce}):
\be Q \int_{\gamma_1} G_{\mu} |\Psi\rangle =
\int_{\gamma_1} \{Q,G_{\mu} \}|\Psi\rangle =
\int_{\gamma_1} P_{\mu} |\Psi\rangle =0.
\label{diezseis}
\ee Similarly, one may construct other invariants tensoring $n$ operators
$G_\mu$ and integrating over $n$-cycles $\gamma_n$:
\be
\int_{\gamma_n} G_{\mu_1}G_{\mu_2}...G_{\mu_n} |\Psi\rangle.
\label{diezsiete}
\ee Notice that since the operator $G_\mu$ is odd and its algebra is  
Poincare-like the integrand in this expression is an $n$-form. It is  
straightforward to prove that these states also satisfy  condition
(\ref{catorce}). Therefore, starting  from a state  $|\Psi\rangle\in \ker
\, Q$ we have built a set of partners or descendants giving rise to a
topological multiplet. The members of a multiplet have well defined
{\it ghost} number. If one assigns ghost number $-1$ to the operator $
G_{\mu}$ the state in (\ref{diezsiete})  has ghost number $-n$ plus the
ghost number of
$|\Psi\rangle$. Of course, $n$
 is bounded by the dimension of the manifold $X$. Among the states
constructed in this way there may be many which are related via another
state which is
$Q$-exact, \ie, which can be written as $Q$  acting on some other state.
Let us try to single out representatives at each level of ghost number in
a given topological multiplet. 

Consider an $(n-1)$-cycle which is the boundary of an $n$-dimensional
surface,
$\gamma_{n-1}=\partial S_n$. If one  builds a state taking such a cycle one
finds ($P_\mu=-i\partial_\mu$),
\be
\int_{\gamma_{n-1}} G_{\mu_1}G_{\mu_2}...G_{\mu_{n-1}} |\Psi\rangle=i
\int_{S_n} P_{[\mu_1} G_{\mu_2}G_{\mu_3}...G_{\mu_{n}]}|\Psi\rangle= iQ
\int_{S_n} G_{\mu_1} G_{\mu_2}...G_{\mu_{n}}|\Psi\rangle,
\label{diezocho}
\ee
\ie, it is $Q$-exact. The symbols [ ] in (\ref{diezocho}) denote that all
indices between  them must by antisymmetrized. In (\ref{diezocho}) use has
been made of (\ref{trece}). This result tells us that  the representatives
we are looking for are built out of the homology cycles of the manifold
$X$. Given a manifold $X$, the homology cycles are equivalence classes
among cycles, the equivalence relation being that two $n$-cycles are
equivalent if they differ by a cycle which is the boundary of an $n+1$
surface. Thus, knowledge on the homology of the manifold on which the TQFT
is defined allows us to classify the representatives among the operators 
(\ref{diezsiete}). Let us assume that
$X$ has dimension $d$ and that its homology cycles are
$\gamma_{i_n}$, $i_n=1,...,d_n$, $n=0,...,d$, being $d_n$ the  dimension
of the 
$n$-homology group, and $d$ the dimension of $X$. Then, the non-trivial
partners or descendants of a given $|\Psi\rangle$ {\it highest-ghost-number
state}  are labeled in the following way:
 \be
\int_{\gamma_{i_n}} G_{\mu_1}G_{\mu_2}...G_{\mu_n} |\Psi\rangle,
\,\,\,\,\,\,\,\,\,i_n=1,...,d_n,\,\,\,\,\,\,\, n=0,...,d.
\label{dieznueve}
\ee

A similar construction to the one just described can be made for  fields.
Starting with a field $\phi(x)$ which satisfies, 
\be [Q,\phi(x)]=0,
\label{veinte}
\ee one can construct other fields using the operators $G_{\mu}$. These
fields, which we will call partners or descendants are antisymmetric
tensors defined as,
\be
\phi^{(n)}_{\mu_1\mu_2...\mu_n}(x)={1\over
n!}[G_{\mu_1},[G_{\mu_2}...[G_{\mu_n},\phi(x)\}...\}\},
\,\,\,\,\,\,\,\, n=1,...,d.
\label{veintep}  
\ee  Using (\ref{trece}) and (\ref{veinte}) one finds that these fields  
satisfy the so-called {\it topological descent equations}:
\be d \phi^{(n)} = i [Q,\phi^{(n+1)}\},
\label{vseis}
\ee where the subindices of the forms have been suppressed for
simplicity,  and the highest-ghost-number field $\phi(x)$ has been denoted
as $\phi^{(0)}(x)$. These equations enclose all the relevant properties of
the observables which are constructed out of  them. They  constitute a
very useful tool to build the observables of the theory. Let us consider
an $n$-cycle and the following quantity:
\be W^{(\gamma_n)}_\phi = \int_{\gamma_n} \phi^{(n)}.
\label{vsiete}
\ee The subindex of this quantity denotes the highest-ghost-number field
out of which the form $\phi^{(n)}$ is generated. The superindex  denotes
the order of such a form as well as the cycle which is utilized in the
integration. Using the topological descent equations (\ref{vseis}) it  is
immediate to prove that
$W^{(\gamma_n)}_\phi$ is indeed an observable:
\be [Q, W^{(\gamma_n)}_\phi\} =\int_{\gamma_n} [Q,\phi^{(n)}\}
=-i\int_{\gamma_n} d\phi^{(n-1)}=0.
\label{vocho}
\ee Furthermore, if $\gamma_n$ is a trivial homology cycle,  
$\gamma_n=\partial S_{n+1}$, one obtains that $W^{(\gamma_n)}_\phi$ is
$Q$-exact,
\be W^{(\gamma_n)}_\phi = \int_{\gamma_n} \phi^{(n)}=
\int_{S_{n+1}} d \phi^{(n)} = i \int_{S_{n+1}} [Q, \phi^{(n+1)}\}
=i[Q,\int_{S_{n+1}}\phi^{(n+1)}\},
\label{vnueve}
\ee and therefore its presence in a vev makes it vanish. Thus, similarly 
to the previous analysis leading to (\ref{dieznueve}), the observables of 
the theory are operators of the form (\ref{vsiete}):
\be W^{(\gamma_{i_n})}_\phi,\,\,\,\,\,\, i_n=1,...,d_n,\,\,\,\,\,\,  
n=0,...,d,
\label{treinta}
\ee where, as before, $d_n$ denotes the dimension of the $n$-homology 
group. Of course, these observables are a basis of observables but one
can  make arbitrary products of them leading to new ones.

One may wonder at this point how it is possible that there may be
observables which depend on the space-time position $x$ and  nevertheless
lead to topological invariants. For example, an observable containing the
zero form
$\phi^{(0)}(x)$ seems to lead to vacuum expectation values which depend  on
$x$ since the space-time position $x$ is not integrated over. A closer
analysis, however, shows that this is not the case. As follows from the
topological descent equation (\ref{vseis}), the derivative of  
$\phi^{(0)}(x)$ with  respect to $x$ is $Q$-exact and therefore such a
vacuum expectation  value is actually independent of the space-time
position.  

The structure of observables described here is common to all cohomological
TQFTs. In these lectures we will review the cases of topological sigma
models, and Donaldson-Witten theory and its generalizations. In the first
case the highest-ghost-number observables are built out of the cohomology
of the target manifold of the sigma model. In the second case they are
obtained from the independent Casimirs of the gauge group under
consideration. Once the highest-ghost-number observables are identified,
their families are constructed solving the topological descent equations
(\ref{vseis}).

We will finish this section pointing out that there is a special set \cite{btuno,dij,australia}, of cohomological TQFTs which play an important role when trying to analyze Euler characters of some moduli spaces. The feature which characterizes these theories is that they possess two topological symmetries. One of the most interesting examples of this kind of theories is a twisted version of $N=4$ supersymmetric Yang-Mills, which has been used recently to carry out a test of $S$-duality in four dimensions.


\vfill
\newpage

\section{The Mathai-Quillen formalism}
\setcounter{equation}{0}

In the rest of these lectures we will restrict ourselves to cohomological
TQFTs. These theories can be constructed from  supersymmetric theories. In
fact, the first examples of TQFTs of this type, four-dimensional
Donaldson-Witten theory \cite{tqft} and two-dimensional topological sigma
models \cite{tsm}, were constructed starting from four-dimensional $N=2$
supersymmetric gauge theory  and two-dimensional
$N=2$ supersymmetric sigma models, respectively. We will discuss their
origin from supersymmetric  theories in a forthcoming section. In this
section we will introduce them within a more mathematical framework, the
Mathai-Quillen formalism, which we first discuss in its simplest form. We
will follow a presentation similar to the one by Blau and
Thompson \cite{blau}.

TQFTs of cohomological type are  characterized by three basic data: fields,
symmetries, and  equations \cite{coho,thompson,moore}. The starting point
is a configuration space ${\cal X}$, whose elements  are fields $\phi_i$
defined on some Riemannian manifold $X$. These fields  are generally acted
on by some group
${\cal G}$ of local transformations  (gauge symmetries, or a diffeomorphism
group, among others), so one is  naturally led to consider the quotient
space
${\cal X}/{\cal G}$. Within this quotient  space, a certain subset or {\it
moduli space}, ${\cal M}$, is singled out by a set of  equations
$s(\phi_i)=0$:
\be  {\cal M}=\{\phi_i\in{\cal X}|s(\phi_i)=0\}/{\cal G}.
\label{moduli}
\ee Within this framework, the topological symmetry $\delta$ furnishes a
representation of the ${\cal G}$-equivariant cohomology on the field 
space. When
${\cal G}$ is the trivial group, $\delta$ is nothing but the de Rham 
operator on the field space. 

The next step consists of building the topological theory associated to 
this moduli problem. We will do this within the framework of the 
Mathai-Quillen formalism \cite{mathai}. This formalism is the most
geometric one  among all the approaches leading to the construction of
TQFTs. It can be applied to  any Witten-type theory. It was first
implemented in the context of TQFT by Atiyah and Jeffrey \cite{jeffrey},
and later further developed in a series of works \cite{blaumathai,moore}.
The basic idea behind this formalism is the  extension to the
infinite-dimensional case of ordinary finite-dimensional geometric
constructions. Soon after the formulation of the first TQFTs it became 
clear that the partition function of these theories was related to the
Euler class of a certain bundle associated to the space of solutions of
the basic equations of the theory. In the finite-dimensional case there
are many different, though equivalent, forms of thinking on the Euler
class, which we will recall bellow. The  Mathai-Quillen formalism
basically consists of generalizing one of these forms to the
infinite-dimensional case. In what follows we will give a brief account on
the fundamentals of the construction. For further details, we refer the 
reader to
 \cite{moore,blau}, where excellent reviews on this approach are presented. 

\subsection{Finite-dimensional case}

Let $X$ be an orientable, boundaryless, compact $n$-dimensional  manifold.
Let us consider an orientable vector bundle ${\cal E}\to X$ of rank
${\hbox{\rm rk}}({\cal E})=2m\leq n$ over $X$. For completeness we  recall
that a vector bundle ${\cal E}$, with  a $2m$-dimensional vector space
$F$ as fibre, over a base manifold $X$, is a topological space with a  
continuous projection, $\pi:{\cal E}\to X$, such that,
$\forall x\in X$, $\exists U_x\subset X$, open set, $x\in U_x$, ${\cal E}$
is a product space, $U_x\times F$, when restricted to $U_x$. This  means
that there exists a homeomorphism $\varphi:U_x\times F\to \pi^{-1}(U)$
which  preserves the fibres, \ie, $\pi(\varphi(x,f))=x$, with $f\in F$.  

There exist two complementary ways of defining the Euler class of  
${\cal E}$,
$e({\cale})\in H^{2m}(X)$:
\begin{enumerate}
\item In terms of sections. A section $s$ of ${\cal E}$ is a map
$s:X\to {\cal E}$ such that $\pi(s(x))=x$. A {\it generic} section is  one
which  is transverse to the zero section, and which therefore vanishes on
a set of dimension $n-2m$. In this context 
$e({\cale})$ shows up as the Poincare dual (in $X$) of the homology  class
defined by the zero locus of a generic section of ${\cale}$.  
\item  In terms of characteristic classes. The approach makes use of  the 
Chern-Weil theory, and gives a representative 
$e_{\nabla}({\cale})$ of $e({\cal E})$ associated to a connection  
$\nabla$ in
${\cale}$, 
\be e_{\nabla}({\cale})=(2\pi)^{-m}\pff (\Omega_{\nabla}),
\label{treintauno}
\ee where $\pff(\Omega_{\nabla})$ stands for the Pfaffian of the
curvature  
$\Omega_{\nabla}$, which is an antisymmetric matrix of two-forms. The
representative
$e_{\nabla}({\cale})$ can be written in ``field-theoretical" form:
\be e_{\nabla}({\cale})=(2\pi)^{-m}\int  
d\chi\ex^{\half\chi_a\Omega^{ab}_{\nabla}\chi_b},
\label{treintados}
\ee by means of a set of real Grassmann-odd variables $\chi_a$,
$a=1,\ldots ,2m$, satisfying the Berezin rules of integration:
\be
\int d\chi_a \chi_b = \delta_{ab}.
\label{treintatres}
\ee
\end{enumerate}

If ${\hbox{\rm rk}}({\cale})=2m=n={\hbox{\rm dim}}(X)$, one can  evaluate
$e({\cal E})$ on $X$ to obtain the Euler number of ${\cal E}$ in two 
different ways:  
\bea
\chi({\cale})&=&\sum_{x_k:s(x_k)=0} (\pm1),\nonumber \\
\chi({\cale})&=& \intl_X e_{\nabla}({\cale}).\nonumber \\ 
\label{treintacuatro}
\eea In the first case, one counts signs at the zeroes of a generic
section. In the second case, an integration of the differential form
(\ref{treintauno}) is performed. Of course, both results coincide, and do
not depend either on the section $s$ (as long as it is generic) or on the
connection $\nabla$.  When
$2m<n$ one can evaluate $e({\cal E})$ on $2m$-cycles or  equivalently take
the product with elements of $H^{n-2m}(X)$ and evaluate it on $X$.  

In the particular case that  ${\cale} \equiv TX$ the expression
$\chi({\cale})=\sum_{x_k:s(x_k)=0} (\pm1)$, which gives the Euler  number
of the base manifold $X$,  can be generalized to a non-generic vector 
field $V$ (which is a section of the tangent bundle),
\be
\chi(X)=\chi(X_V),
\label{treintacinco}
\ee where $X_V$ is the zero locus of $V$, which is not necessarily  
zero-dimensional.

In this framework the Mathai-Quillen formalism gives a representative of 
the Euler class, $e_{s,\nabla}({\cale})$, which interpolates between the 
two approaches sketched above. It depends explicitly on both, a section,
$s$, and a connection, $\nabla$, on ${\cal  E}$:
\bea e_{s,\nabla}({\cale})&\in&[e({\cale})],\nonumber\\
\chi({\cale})&=&\intl_X e_{s,\nabla}({\cale}),\qquad  ({\rm if}\,\, 2m=n).
\label{treintaseis}
\eea The construction of $e_{s,\nabla}$ is given by the formalism. First,
it provides an explicit representative of the Thom class \cite{tu},
$\Phi({\cal E})$, of ${\cal E}$. Let ${\cal E}\to X$ be a vector bundle of
rank
$2m$ with  fibre $F$, and let us consider the cohomology of forms with
Gaussian decay along  the fibre. By integrating the form along the fibre
one has an explicit isomorphism (the Thom isomorphism) between $k$ forms
over ${\cal E}$  and $k-2m$ forms over $X$. This isomorphism can be  made
explicit with the  aid of the Thom class, whose representative,
$\Phi({\cal E})$, is a closed 
$2m$-form over ${\cal E}$ with Gaussian decay along the fibre such that 
its integral over the fibre is unity. In terms of this form, and given any
arbitrary $p$-form $\omega$ over $X$, its image under the Thom isomorphism
is the $p+2m$ form $\pi^*(\omega)\wedge\Phi({\cal E})$, which by
construction has Gaussian decay along the fibre. $\pi^*(\omega)$ is the
pull-back of
$\omega$ by  the projection $\pi:{\cal E}\to X$. If $s$ is any section of
${\cal E}$, the pull-back of the Thom form under $s$, $s^*\Phi({\cal E})$,
is a  closed form in the same cohomology class as the Euler class $e({\cal
E})$. If
$s$ is a  generic section, then $s^{*}\Phi({\cal E})$ is the Poincare
dual  of the zero locus of $s$. Mathai and Quillen constructed an explicit
representative, $\Phi_{\nabla}({\cale})$, of the Thom form in terms of  a
connection $\nabla$ in ${\cale}$. Its pullback by a section $s$,
$e_{s,\nabla}({\cale})=s^{*}\Phi_{\nabla}({\cale})$, is  represented as a
Grassmann integral:
\be e_{s,\nabla}({\cale})=(2\pi)^{-m}\int d\chi\ex^{-\half \vert
s\vert^2+\half\chi_a\Omega^{ab}_{\nabla}
\chi_b+i\nabla s^a\chi_a}.
\label{treintasiete}
\ee As a consistency check note that, as follows from (\ref{treintados}),  
$e_{s=0,\nabla}({\cale})=e_{\nabla}({\cale})$,
\ie, the pull-back of the Mathai-Quillen representative by the zero 
section gives back the Euler class of ${\cale}$. $e_{s,\nabla}({\cale})$
is a closed
$2m$-form. This can be verified after integrating over the Grassmann-odd
variables  
$\chi_a$. It is closed because the exponent is invariant under the
transformations:
\be
\delta s=\nabla s,\qquad \delta \chi_a=is_a.
\label{treintaocho}
\ee

It is possible to find a nice physics-like form for  
$e_{s,\nabla}({\cale})$. To this end we introduce grassmann odd real
variables
$\psi^\mu$ with the correspondence:
\be  dx^\mu\leftrightarrow\psi^\mu,
\label{treintanueve}
\ee
\be
\omega={1\over  {p!}}\omega_{\mu_1\cdots\mu_p}dx^{\mu_1}\wedge\cdots\wedge
dx^{\mu_p}\leftrightarrow 
\omega(\psi)={1\over
{p!}}\omega_{\mu_1\cdots\mu_p}\psi^{\mu_1}\cdots\psi^{\mu_p}.
\label{cuarenta}
\ee The integral over $X$ of a top-form, $\omega^{(n)}$, is therefore
given by a simultaneous conventional integration over $X$ and a Berezin 
integration over the $\psi$'s:
\be
\intl_X \omega^{(n)}=\int _X dx \int d\psi \, \omega^{(n)}(\psi).
\label{cuarentauno}
\ee In this language, the Mathai-Quillen representative 
(\ref{treintasiete}) can be rewritten as:
\be e_{s,\nabla}({\cale})(\psi)=(2\pi)^{-m}\int d\chi\ex^{-\half \vert
s\vert^2+\half\chi_a\Omega^{ab}_{\nabla}(\psi)
\chi_b+i\nabla s^a(\psi)\chi_a},
\label{cuarentados}
\ee and, for example, in the case $n=2m$, one has the following
expression  for the Euler number of $\cale$: 
\be
\chi({\cale})=(2\pi)^{-m} \intl_X dx  d\psi d\chi \ex^{-\half \vert
s\vert^2+\half\chi_a\Omega^{ab}_{\nabla}(\psi)
\chi_b+i\nabla s^a(\psi)\chi_a}.
\label{cuarentatres}
\ee It is worth to remark that (\ref{cuarentatres}) looks like the 
partition function of a field theory whose ``action" is: 
\be A(x,\psi,\chi)=
\half \vert s\vert^2-\half\chi_a\Omega^{ab}_{\nabla}(\psi)
\chi_b-i\nabla s^a(\psi)\chi_a.
\label{cuarentacuatro}
\ee This action is invariant under the transformations: 
\be
\delta x^\mu=\psi^\mu,\qquad \delta\psi^\mu=0,\qquad  
\delta\chi_a=is_a.
\label{cuarentacinco}
\ee

We mentioned above that the Mathai-Quillen representative interpolates
between the two different approaches to the Euler class of a vector
bundle. This statement can be made more precise as follows. The
construction of
$e_{s,\nabla}({\cale})$  is such that it is cohomologous to
$e_{\nabla}({\cale})$ for any choice of a generic section $s$. Take for  
example the case $n=2m$, and rescale $s\to\gamma s$. Nothing should
change, so in  particular:
\be 
\chi({\cale})=\intl_X e_{\gamma s,\nabla}({\cale}).
\label{cuarentaseis}
\ee We can now study (\ref{cuarentaseis}) in two different limits:
\begin{enumerate}

\item Limit $\gamma\to 0$: after using (\ref{treintados}),
$\chi({\cale})=(2\pi)^{-m}\int\pff (\Omega_{\nabla})$.
\item Limit $\gamma\to\infty$: the curvature term in  (\ref{cuarentatres})
can be neglected, leading to $\chi({\cale})=\sum_{x_k:s(x_k)=0}(\pm 1)$.
These signs are generated by the ratio of the determinants of $\nabla s$
and its modulus, which result from the Gaussian integrations after
expanding around each zero
$x_k$.
\end{enumerate}

\noindent Hence, we recover from this unified point of view the two
complementary  ways to define the Euler class described at the beginning
of the section. 

Let us work out an explicit example. To be definite we will consider
${\cale}=TX$. The section $s$ is taken as a vector field $V$ on  
$X$, which we assume to be generic.  The action (\ref{cuarentacuatro}),
after rescaling $V\to\gamma V$, takes the form: 
\be  A(x,\psi,\chi)=
 \half\gamma^2 g_{\mu\nu}V^\mu V^\nu-{1\over4}\chi_a
R^{ab}{}_{\mu\nu}\psi^\mu\psi^\nu
\chi_b-i\gamma\nabla_\mu V^\nu \psi^\mu e^a_\nu \chi_a.
\label{cuarentaocho}
\ee To compute (\ref{cuarentatres}) in the limit $\gamma\to\infty$ we
expand around the zeroes, $x_k$, of $V$ ($V^\mu(x_k)=0$):
\be
\chi=\sum_{x_k}\intl_X dx d\psi d\chi(2\pi)^{-m}\ex^{
 -\half\gamma^2 g_{\mu\nu}\partial_\sigma V^\mu\partial_\rho
 V^\nu x^\sigma x^\rho+{1\over4}\chi_a  R^{ab}{}_{\mu\nu}\psi^\mu\psi^\nu
\chi_b +i\gamma\partial_\mu V^\nu \psi^\mu e^a_\nu \chi_a}.
\label{cuarentanueve}
\ee Next, we rescale the variables in the following way: 
\begin{eqnarray} x&\to & \gamma^{-1}x,\nonumber\\ dx&\to & \gamma^{-1}dx,
\nonumber\\  
\psi&\to&\gamma^{-\half}\psi,\nonumber\\  d\psi &\to & \gamma^{\half}d\psi,
\\
\chi &\to& \gamma^{-\half}\chi,\nonumber\\ d\chi &\to &
\gamma^{\half}d\chi. 
\nonumber
\label{cincuenta}
\end{eqnarray} Notice that the measure is invariant under this rescaling. 
Using the shorthand notation for the Hessian, $H^{(k)\mu}_\sigma
=\partial_\sigma V^\mu \big |_{x_k}$,  one finds, after taking the limit
$\gamma\to\infty$:
\bea  &&\chi = \sum_{x_k}\intl_X dx d\psi d\chi (2\pi)^{-m}\ex^{ -\half
g_{\mu\nu}H^{(k)\mu}_\sigma H^{(k)\nu}_\rho  x^\sigma x^\rho
+iH_{\mu}^{(k)\nu}
\psi^\mu e^a_\nu
\chi_a}\nonumber\\ &&
=\sum_{x_k}(2\pi)^{-m}{{(\sqrt{2\pi})^{2m}}\over{\sqrt{g}\vert  {\hbox{\rm
det}} ~H^{(k)}\vert}}{e\over{(2m)!}}(2m)!{\hbox{\rm det}}~H^{(k)}
=\sum_{x_k}{{\hbox{\rm det}}~H^{(k)}
\over{\vert {\hbox{\rm det}}~H^{(k)}\vert}},
\label{cincuentauno}
\eea which indeed corresponds to the Euler number of $X$ in virtue of the
Poincare-Hopf theorem.

It is possible to introduce auxiliary fields in the formulation. In  the
example under consideration, after using,
\be
\ex^{-\half\gamma^2 g_{\mu\nu}V^\mu V^\nu}={\gamma^{2m}\sqrt{g}
\over{ (2\pi)^{m}}}\int dB \ex^{-\half\gamma^2 (g_{\mu\nu}B^\mu B^\nu+  
2iB_\nu V^\nu)},
\label{cincuentatres}
\ee  
 being $B^\mu$ an auxiliary field, the Euler number resulting from
(\ref{cuarentaocho}) can be rewritten as: 
\be 
\chi=\intl_X dx  d\psi d\chi dB  {\gamma^{2m}\sqrt{g}\over{(2\pi)^{2m}}}
\ex^{-\half\gamma^2 (g_{\mu\nu}B^\mu B^\nu+2iB_\nu V^\nu) 
+{1\over4}\chi_a R^{ab}{}_{\mu\nu}\psi^\mu\psi^\nu
\chi_b +i\gamma\nabla_\mu V^\nu \psi^\mu e^a_\nu \chi_a}.
\label{cincuentacuatro}
\ee  Making the redefinitions: 
\bea
\psi^\mu&\to & \gamma^{\half}\psi^\mu, \nonumber\\ d\psi&\to &
(\gamma^{-\half})^{2m} d\psi,\nonumber\\
\chi_a &\to & \gamma^{\half}e_{a\mu}\bar\psi^\mu, \\  d\chi  &\to&
(\gamma^{-\half})^{2m} {1\over\sqrt{g}}d\bar\psi,\nonumber
\label{cincuentacinco}
\eea one obtains: 
\be 
\chi=\intl_X dx d\psi d\bar\psi dB{1\over{(2\pi)^{2m}}}
\ex^{-\gamma^2 \bigl(\half g_{\mu\nu}(B^\mu B^\nu+2iB^\mu V^\nu)  
-{1\over4}R^{\rho\sigma}{}_{\mu\nu}\bar\psi_\rho\bar\psi_\sigma
\psi^\mu\psi^\nu
 -i\nabla_\mu V^\nu \psi^\mu \bar\psi_\nu\bigr)}.
\label{cincuentaseis}
\ee  This looks like the partition function of a topological quantum field
theory, in which  
$g = 1/\gamma$ plays the role of the coupling constant. Furthermore, the
exponent of (\ref{cincuentaseis}) is invariant under  the symmetry:
\bea
\delta x^\mu&=&\psi^\mu,\nonumber\\
\delta\psi^\mu&=&0,\nonumber\\
\delta\bar\psi_\mu&=&B_\mu,\\
\delta B_\mu&=& 0.\nonumber
\label{cincuentasiete}
\eea Notice that $\delta^2=0$. In fact, one easily finds that the exponent
is indeed $\delta$-exact:
\be 
\chi=\intl_X dx d\psi d\bar\psi dB{1\over{(2\pi)^{2m}}}
\ex^{-\gamma^2\delta \bigl(\half \bar\psi_\mu(B^\mu +2iV^\mu  
+\Gamma^\sigma_{\tau\nu}\bar\psi_\sigma\psi^\nu g^{\mu\tau}).
\bigr)}.
\label{cincuentaocho}
\ee This result makes possible to use field-theoretical arguments to
conclude that $\chi$ is independent of the coupling $\gamma$ and of the 
metric
$g_{\mu\nu}$. 

The Mathai-Quillen formalism can be recasted in a conventional BRST 
language in which $\Psi=\half \bar\psi_\mu(B^\mu +2iV^\mu  
+\Gamma^\sigma_{\tau\nu}\bar\psi_\sigma\psi^\nu g^{\mu\tau})$ plays the 
role of a {\it gauge fermion}, and the exponent of (\ref{cincuentaocho})
can be regarded as an action
$A=\delta\Psi$.  In many situations one has the general pattern: 
\be
\Psi=\chi_a\bigl(\,\underbrace{s^a}_{{\scriptstyle {\rm section}}}+
\underbrace{\theta^{ab}\chi_b}_{{\scriptstyle {\rm connection}}}+
\underbrace{B^a}_{{\scriptstyle {\rm auxiliary}}}\bigr).
\label{cincuentanueve}
\ee

In order to fully understand the construction, let us be more specific
with our example. Consider the two-sphere $S^2$ with the standard
parametrization: 
\bea
\alpha :(0,\pi)\times(0,2\pi)&\too& {\bf R}^3,\nonumber\\
(\theta,\varphi)\,\,\,\,\,\,\,\,\,\,\,\,\,\,
&\too&(\sin\theta\cos\varphi,\sin\theta\sin
\varphi,\cos\theta).
\label{sesenta}
\eea In terms of these coordinates, we have the relations:
\be ds^2=d\theta^2+\sin^2\theta d\varphi^2,\qquad 
g_{\mu\nu}=\pmatrix{1&0\cr 0&\sin^2\theta\cr},
\label{sesentauno}
\ee and the following values for the Christoffel symbol 
($\Gamma^\lambda_{\mu\nu}=\half 
g^{\lambda\sigma}(\partial_{(\mu}g_{\nu)\sigma} -\partial_\sigma
g_{\mu\nu})$),
\bea
\Gamma^\theta_{\theta\theta}&=&\Gamma^\varphi_{\varphi\varphi}=
\Gamma^\varphi_{\theta\theta}=  
\Gamma^\theta_{\theta\varphi}=0,\nonumber\\
\Gamma^\theta_{\varphi\varphi}&=&-\sin\theta\cos\theta,\quad 
\Gamma^\varphi_{\theta\varphi}={{\cos\theta}\over{\sin\theta}}.
\label{sesentatres}
\eea Let us pick an orthonormal frame:
\be  e_a^\mu=\pmatrix{1&0\cr 0&{1\over{\sin\theta}}\cr},\qquad  e^a_\mu=
\pmatrix{1&0\cr 0&\sin\theta\cr},
\label{sasentacuatro}
\ee where the vielbeins satisfy the standard relations: $e_a^\mu e_b^\nu
g_{\mu\nu}=\delta_{ab}$,
$e^a_\mu e^b_\nu
\delta_{ab}= g_{\mu\nu}$. The Riemann curvature tensor
($R^\lambda{}_{\mu\nu\kappa}=\partial_{[\kappa}\Gamma^\lambda_{\nu]\mu}+  
\Gamma^\tau_{\mu[\nu}\Gamma^\lambda_{\kappa]\tau}$) in $(\theta,\varphi)$
coordinates is given  by:
\be  R^\theta{}_{\varphi\varphi\theta}=\sin^2\theta,
\label{sesentacuatro}
\ee while the curvature two-form $\Omega^{ab}$ takes the form: 
\be 
\Omega^{12}=R^{12}{}_{\varphi\theta} d\varphi \wedge d\theta =  e^1_\theta 
e^2_\varphi g^{\varphi\varphi} R^\theta{}_{\varphi\varphi\theta} d\varphi
\wedge d\theta =\sin\theta d\varphi \wedge d\theta. 
\label{sesentacinco}
\ee

Next let us consider the vector field\footnote{This vector field is
actually equivalent to the one considered in  \cite{blau} but with a better
choice of coordinates for our purposes.}\,\,:
\be V^a=(\sin\varphi,\cos\varphi\cos\theta)\to
V^\mu=(\sin\varphi,\cos\varphi\cot\theta).
\label{leman}
\ee  This vector field has zeroes at 
$\varphi=0,~\theta={\pi\over{2}}$ and  
$\varphi=\pi,~\theta={\pi\over{2}}$. The components of the form $\nabla
V^a$ are:
\be 
\nabla_\theta V^\varphi =-\cos\varphi,\quad 
\nabla_\varphi V^\theta =\sin^2\theta\cos\varphi,\quad 
\nabla_\theta V^\theta=\nabla_\varphi V^\varphi=0,
\label{setentacuatro}
\ee or, alternatively, 
\be
\nabla_\theta V^a =e^a_\mu\nabla_\theta V^\mu
=(0,-\cos\varphi\sin\theta),\qquad 
\nabla_\varphi V^a =e^a_\mu\nabla_\varphi V^\mu 
=(\cos\varphi\sin\theta,0),
\label{setentacinco}
\ee and therefore,
\be 
\nabla V^a  =(\sin^2\theta\cos\varphi d\varphi,-\cos\varphi\sin\theta  
d\theta).
\label{setentaseis}
\ee

The Euler class representative,
\be  e_{V,\nabla}(TS^2)={1\over {2\pi}}\int d\chi_1 d\chi_2\ex^{ -\half
V^aV^a+\half\chi_a\Omega^{ab}_{\nabla}\chi_b+i\nabla V^a  
\chi_a},
\label{sesentanueve}
\ee after performing the rescaling $V^a\to\gamma V^a$,
\bea && -\half V^aV^a+\half\chi_a \Omega^{ab}_{\nabla}\chi_b+i\nabla  V^a
\chi_a\too
\nonumber\\ && -{\gamma^2\over{2}}
(\sin^2\theta+\cos^2\varphi\cos^2\theta)-\chi_1\chi_2\sin\theta 
d\theta\wedge d\varphi
\nonumber\\ &&\,\,\,\,\,\,\,\,\,\,\,\,\,\,\,\,\,\,\,\, + i\gamma
(\sin^2\theta\cos\varphi d\varphi\chi_1-\cos\varphi\sin\theta
d\theta\chi_2),
\label{setentasiete}
\eea  becomes:
\be e_{V,\nabla}(TS^2)={1\over
{2\pi}}\ex^{-{\gamma^2\over{2}}(\sin^2\theta+\cos^2\varphi\cos^2\theta)} 
\sin\theta (1+\gamma^2\cos^2\varphi\sin^2\theta)d\theta\wedge d\varphi.
\label{setentaocho}
\ee

The Euler number of $S^2$ is given by the integral:
\be
\chi(S^2)={1\over 
{2\pi}}\intl_0^{2\pi}d\varphi\intl_0^{\pi}d\theta\sin\theta 
\ex^{-{\gamma^2\over{2}}(\sin^2\theta+\cos^2\varphi\cos^2\theta)}
(1+\gamma^2\cos^2\varphi\sin^2\theta).
\label{setentanueve}
\ee  Although $\gamma$ appears explicitly in this expression, the result
of the integration should be independent of $\gamma$.  The reader is urged
to prove it (we do not know of any analytical proof, we have only
numerical evidence). One can perform, however, two independent checks. On
the one hand, in the limit
$\gamma\to 0$, (\ref{setentanueve}) gives trivially the correct  result,
$\chi(S^2)=2$. On the other hand, one can explore the opposite limit,
$\gamma\to\infty$, where the integral,
\be 
\chi(S^2)=\intl_{S^2}dx\int d\psi^1d\psi^2\int d\chi_1 d\chi_2 
{1\over{2\pi}}\ex^{-A(x,\psi,\chi)},
\label{ochenta}
\ee  with $A(x,\psi,\chi)=\half
V^aV^a-\half\chi_a\Omega^{ab}_{\nabla}\chi_b-i\nabla V^a \chi_a$, is 
dominated by the zeroes of $V^a$.  We expand around them:

\vspace{1pc}  (a) $\theta={\pi\over{2}}+x$, $\varphi=0+y$.  We get:
\bea 
V^aV^a=\sin^2\varphi+\cos^2\varphi\cos^2\theta&=&x^2+y^2+\cdots\nonumber 
\\ {1\over{4}}\chi_a R^{ab}{}_{\mu\nu}\psi^\mu\psi^\nu\chi_b=
\chi_1\chi_2\psi^\theta\psi^\varphi R^{12}{}_{\theta\varphi}&=&
\sin\theta\chi_1\chi_2\psi^\theta\psi^\varphi=(1-{x^2\over{2}}+\cdots)
\chi_1\chi_2\psi^\theta\psi^\varphi\nonumber\\
\nabla_\mu V^\nu\psi^\mu 
e^a_\nu\chi_a=-\cos\varphi\sin\theta\psi^\theta\chi_2
&+&\cos\varphi\sin^2\theta\psi^\varphi\chi_1=
-\psi^\theta\chi_2+\psi^\varphi\chi_1+\cdots\nonumber\\
\label{ochentauno}
\eea  Next, performing the rescaling: 
\bea  x,y&\to& \gamma^{-1}x,\gamma^{-1} y,\nonumber\\ 
\psi^\mu &\to& \gamma^{-\half}\psi^\mu, \\ 
\chi_a &\to &\gamma^{-\half}\chi_a,\nonumber
\label{ochentados}
\eea we obtain: 
\be {1\over{2\pi}}\int dxdy\,\ex^{-\half(x^2+y^2)}\int d\chi_1d\chi_2
d\psi^\theta d\psi^\varphi  
\ex^{i(-\psi^\theta\chi_2+\psi^\varphi\chi_1)}=1.
\label{ochentatres}
\ee

\vspace{1pc} (b) Similarly, expanding around the second zero, 
$\theta={\pi\over{2}}+x$, $\varphi=\pi+y$, one finds  the contribution:  

\be {1\over{2\pi}}\int dxdy\ex^{-\half(x^2+y^2)}\int d\chi_1d\chi_2
d\psi^\theta d\psi^\varphi  
\ex^{i(\psi^\theta\chi_2-\psi^\varphi\chi_1)}=1.
\label{ochentacuatro}
\ee

\vspace{1pc} \noindent Therefore, in the limit $\gamma\to\infty$ we have
reproduced the behavior described in (\ref{cincuentauno}),
\be
\chi(S^2)=\left( {{{\hbox{\rm det}}\pmatrix{0&-1\cr1&0\cr}}\over 
{\left\vert{\hbox{\rm det}}\pmatrix{0&-1\cr1&0\cr}\right\vert}}+ 
{{{\hbox{\rm det}}\pmatrix{0&1\cr-1&0\cr}}\over {\left\vert{\hbox{\rm
det}}\pmatrix{0&1\cr-1&0\cr}\right\vert}}\right)=2.
\label{ochentacinco}
\ee

\subsection{Infinite-dimensional case}

We now turn into the study of the infinite-dimensional case. The main
complication that one finds in this case is that $e({\cal E})$ is not
defined. By taking advantage of what we have learned so far, we could try
to  use the Mathai-Quillen formalism to define something analogous to an
Euler  class for
${\cale}$. It turns out that this is actually possible. The outcome of  the
construction is what is called a regularized Euler number for the  bundle
${\cale}$. Unfortunately, it depends  explicitly on the  section  chosen
for the construction, so it is important to make good selections.

The outline of the construction is as follows.  First  recall that, as
stated in (\ref{treintacinco}), in the finite-dimensional case
$\chi(X)=\chi(X_V)$ when $V$ is non-generic, \ie, when its zero locus,
$X_V$,  has dimension ${\hbox{\rm dim}}(X_V)<2m$. For $X$ infinite
dimensional the  idea is to introduce a vector field $V$ with
finite-dimensional zero locus.  The regularized Euler number of ${\cale}$
would be then defined as:
\be
\chi_V(X)=\chi(X_V),
\label{ochentaseis}
\ee which explicitly depends on $V$.  By analogy with the
finite-dimensional case one expects that:
\be
\chi_V(X)=\intl_X e_{V,\nabla}(TX),
\label{ochentasiete}
\ee as a functional integral, where $e_{V,\nabla}(TX)$ is meant to be the
Mathai-Quillen representative for the corresponding Euler class. 

In  general, the regularized Euler number $\chi_s({\cale})$ of an
infinite-dimensional vector bundle ${\cale}$ is given by:
\be
\chi_s({\cale})=\intl_X e_{s,\nabla}({\cale}),
\label{ochentaocho}
\ee where $e_{s,\nabla}({\cale})$ is given by the Mathai-Quillen
formalism. The construction of $e_{s,\nabla}({\cale})$ will be illustrated
by the description of several examples. This construction follows the
pattern of the finite dimensional case. Before entering into the
discussions of these examples it is important to remark that equation
(\ref{ochentaocho}) makes sense when the zero locus of  $s$, $X_s$, is
finite dimensional.
$\chi_s({\cale})$ is the Euler number of  some finite-dimensional vector
bundle over $X_s$, and it corresponds to the regularized Euler number of
the infinite-dimensional bundle ${\cale}$. Of course, $\chi_s({\cale})$
depends on
$s$, but if $s$ is naturally associated to ${\cale}$ one expects to obtain
interesting topological information.

\subsubsection{Supersymmetric quantum mechanics}

Let $X$ be a smooth, orientable, Riemannian manifold with metric  
$g_{\mu\nu}$. The loop space, $LX$, is defined by the set of smooth maps:
\bea  x:S^1&\to& X,\nonumber\\ t\in[0,1]&\to& x^\mu(t),\\ x^\mu(0)&=&
x^\mu(1).\nonumber
\label{ochentanueve}
\eea Let us denote by $T(LX)$ the tangent vector bundle to $LX$, with  
fibre
${\cal F}=T_x(LX)=\Gamma(x^*(TX))$. A  vector field over $LX$ has the form:
\be  V(x)=\oint dt V^\mu(x(t)){\partial\over{\partial x^\mu(t)}}.
\label{noventa}
\ee The metric on $X$ provides a natural metric for $T_x(LX)$: let $V_1,~
V_2\in T_x(LX)$, then,
\be {\hat g}_x(V_1,V_2)=\oint dt 
g_{\mu\nu}(x(t))V_1^\mu(x(t))V_2^\nu(x(t)).
\label{noventauno}
\ee The Levi-Civita connection on $LX$ is the pullback connection from  
$X$:
\bea
\nabla V =\oint dt_1 dt_2 \left [ {{\delta V^\mu(x(t_1))}\over{\delta
x^\nu(t_2)}} +  
\Gamma^\mu_{\nu\lambda}(x(t_2))V^\lambda(x(t_1))\delta(t_1-t_2)
\right]\, {\partial\over{\partial x^\mu(t_1)}}\otimes {\hat  d}x^\nu(t_2),
\nonumber\\
\label{noventados}
\eea where $\{\partial/\partial x^\mu(t)\}$ is a basis of $T_x(LX)$,  and  
$\{{\hat d}x^\nu(t)\}$ is a basis of $T^{*}_x(LX)$. Let us consider the 
vector field: 
\be  V^\mu(x)={d\over{dt}} x^\mu \equiv \dot x^\mu.
\label{noventatres}
\ee The zero locus of $V$ is the space of constant loops, $(LX)_V=X$. 
Therefore, the regularized Euler number of $LX$ is the Euler number of  
$X$ itself:
\be
\chi_V(LX)=\chi((LX)_V)=\chi(X).
\label{knie}
\ee

Let us construct the Mathai-Quillen representative for this Euler  number
following the same procedure as in the finite-dimensional case:
\bea
\matrix{{\rm Finite}\cr{\rm Dimensional}\cr{\rm Case}\cr}&&\cases{
\chi=\intl_X dx d\psi d\bar\psi dB{1\over{(2\pi)^{2m}}}
\ex^{-\gamma^2\delta\Psi(x,\psi,\bar\psi,B)},\cr
\Psi(x,\psi,\bar\psi,B)=\half \bar\psi_\mu(B^\mu +2iV^\mu  
+\Gamma^\sigma_{\tau\nu}\bar\psi_\sigma\psi^\nu g^{\mu\tau}),\cr
\delta x^\mu=\psi^\mu,\qquad \delta\psi^\mu=0,\cr
\delta\bar\psi_\mu=B_\mu,\qquad\delta B_\mu = 0.\cr}\nonumber\\
\matrix{{\rm Supersymmetric}\cr{\rm Quantum}\cr{\rm Mechanics}\cr}&&\cases{
\chi_V=\intl_{LX} dx d\psi d\bar\psi dB{1\over{(2\pi)^{2m}}}
\ex^{-\gamma^2\delta\Psi(x,\psi,\bar\psi,B)},\cr
\Psi(x,\psi,\bar\psi,B)=\half \oint dt\bar\psi_\mu(B^\mu +2i\dot x^\mu 
+\Gamma^\sigma_{\tau\nu}\bar\psi_\sigma\psi^\nu g^{\mu\tau}),\cr
\delta x^\mu(t)=\psi^\mu(t),\qquad \delta\psi^\mu(t)=0,\cr
\delta\bar\psi_\mu(t)=B_\mu(t),\qquad\delta B_\mu(t) =  0.\cr}\nonumber\\
\label{noventacuatro}
\eea

In order to evaluate $\chi_V(LX)$ we first integrate out  the auxiliary
fields in the action in (\ref{noventacuatro}). One finds:
\be 
\chi_V=\int {{dx d\psi 
d\bar\psi}\over{(2\pi)^{m}}}{1\over{\sqrt{g}\gamma^{2m}}}
\ex^{ -\gamma^2\oint dt \left[\half g_{\mu\nu}\dot x^\mu \dot x^\nu 
-i\bar\psi\nabla_t \psi-{1\over4}
R^{\rho\sigma}{}_{\mu\nu}\bar\psi_\rho\bar\psi_\sigma\psi^\mu
\psi^\nu\right] },
\label{noventasiete}
\ee  where,
\be
\nabla_t\psi^\mu=\dot\psi^\mu+\Gamma^\mu_{\nu\sigma}\psi^\sigma\dot  x^\nu.
\label{noventaocho}
\ee The action in the exponential of (\ref{noventasiete}) is precisely the
action corresponding to supersymmetric quantum mechanics \cite{morse}. The  
$\delta$-transformations become:
\be
\delta x^\mu=\psi^\mu ,\qquad \delta\psi^\mu=0,\qquad  
\delta\bar\psi_\mu= -ig_{\mu\nu}\dot x^\nu
-\Gamma^\sigma_{\mu\nu}\bar\psi_\sigma\psi^\nu,
\label{noventanueve}
\ee which close only on-shell, \ie, $\delta^2=0$, modulo field equations. 
As discussed below, they can be regarded as supersymmetry
transformations.  

At this point it is convenient to discuss an additional symmetry which  is
present in the systems under consideration: the ghost number symmetry.  
The
$\delta$-symmetry is compatible with the following ghost number 
assignment:
\vspace{1pc}
\begin{center}
\begin{tabular}{||c|c|c|c|c||}
\hline
$\delta$ & $x$ & $\psi$ & $\bar\psi$ & $B$\\
\hline
$1$ & $0$ & $1$ & $-1$ & $0$ \\
\hline
\end{tabular}
\end{center}
\vspace{1pc}
\noindent The action is ghost number invariant, as it is the measure
itself.  In fact,
$\# \psi$-zero modes $=\# \bar\psi$-zero modes, \ie,  
${\hbox{\rm dim}} ({\hbox{\rm ker}}\nabla_t)= {\hbox{\rm dim}} ({\hbox{\rm
coker}}  
\nabla_t)$.
${\hbox{\rm ker}}\nabla_t$ corresponds to the tangent space at a  constant
bosonic mode, \ie, the tangent space at the zero-locus of a section or 
moduli space: if $\dot x^\mu=0$ and $x^\mu\to x^\mu+\delta x^\mu$ then, 
\be {d\over{dt}}\delta x^\mu=0 \Leftrightarrow \nabla_t\delta x^\mu\big  
|_{\dot x^\mu=0}=0.
\label{noventacinco}
\ee Thus ${\hbox{\rm ker}} \nabla_t$ provides the directions in which a 
given bosonic zero-mode can be deformed into a nearby bosonic zero-mode.
The  ghost number symmetry is potentially anomalous. In this case:
\be {\hbox{\rm Ghost number anomaly}}={\hbox{\rm dim}} ({\hbox{\rm
ker}}\nabla_t)- {\hbox{\rm dim}} ({\hbox{\rm coker}}  
\nabla_t)=0,
\label{noventaseis}
\ee but in general it does not vanish.

Let us compute  $\chi_V(LX)$  in the limit $\gamma\to\infty$. In this 
limit the exact result is obtained very simply by considering the
expansion  of the exponential around bosonic and fermionic zero modes:
\begin{itemize}
\item Bosonic part: $\dot x^\mu=0\to x^\mu~{\hbox{\rm constant}}$
\item Fermionic part: $\cases{\psi^\mu(t)=\psi^\mu +{\hbox{\rm  non-zero
modes}}\cr
\bar\psi^\mu(t)=\bar\psi^\mu +{\hbox{\rm non-zero modes}}\cr}$
\end{itemize} The integration over the non-zero modes is trivial since the
$\delta$ symmetry  implies that the ratio of  determinants is equal to
$1$. The integration over the zero modes gives:
\bea 
\chi_V&=&\intl_X dx {(2\pi)^{-m}\over{\gamma^{2m}\sqrt{g}}}
\int \left[ \prod_{\mu=1}^{2m}d\psi^\mu\right]
\left[ \prod_{\nu=1}^{2m} d\bar\psi^\nu\right]
\ex^{ \gamma^2{1\over4}
R^{\rho\sigma}{}_{\mu\nu}\bar\psi_\rho\bar\psi_\sigma\psi^\mu\psi^\nu
}\nonumber\\ &=& \intl_X {1\over{\gamma^{2m}(2\pi)^{m}}}
\int\left( \prod_{a=1}^{2m}d\chi_a \right)
\ex^{ {{\gamma^2}\over2}
\chi_a\Omega^{ab}\chi_b}=\intl_X {1\over{(2\pi)^m}}\pff (\Omega^{ab})
\nonumber\\ &=& \chi(X)
\label{cien}
\eea  where $\chi_a=e^\mu_a\bar\psi_\mu$.

	In the general case, the measure is not ghost-number invariant.  To get a
non-vanishing functional integral one needs to introduce operators with  
non-zero ghost number. Knowledge of the ghost number anomaly gives
information  on the possible topological invariants, \ie, on the possible
non-vanishing  vacuum expectation values of the theory, so in the end it
provides a  selection rule. In the most interesting situations the ghost
number anomaly is  given by the index of some operator in the theory. In
general, the bosonic  zero-modes  provide the zero-locus of the section,
whereas the fermionic zero-modes are related to the possible deformations
of the bosonic zero-modes. 

To finish this quick tour through supersymmetric quantum mechanics, it is
interesting to recall that $\chi_V$ can be computed using Hamiltonian 
methods \cite{constraints,luisdos,friedan}.  The expression
(\ref{noventasiete}) possesses a second $\delta$-like symmetry,
$\bar\delta$:
\be
\begin{array}{cclcccl}  
\delta x^\mu &=& \psi^\mu,& 
\,\,\,\,\,\,\,\,\,\,\,\,\,\,\ & \bar\delta x^\mu &=& \bar\psi^\mu,  
\nonumber\\ {\delta}\psi^\mu &=& 0,&
\,\,\,\,\,\,\,\,\,\,\,\,\,\,\ & 
\bar\delta\bar\psi_\mu &=& 0 , \nonumber\\
\delta\bar\psi_\mu &=& -ig_{\mu\nu}\dot x^\nu
-\Gamma^\sigma_{\mu\nu}\bar\psi_\sigma\psi^\nu, &  
\,\,\,\,\,\,\,\,\,\,\,\,\,\,\ &
\bar\delta\psi_\mu &=& -ig_{\mu\nu}\dot x^\nu
-\Gamma^\sigma_{\mu\nu}\psi_\sigma\bar\psi^\nu,
 \nonumber\\ 
\end{array}
\label{cdos}
\ee One finds, after using the field equations, that:
\be
\delta^2=0,\qquad \bar\delta^2=0,\qquad  
\delta\bar\delta+\bar\delta\delta= {d\over{dt}},
\label{ctres}
\ee which, in terms of operators,
\be
\delta\leftrightarrow Q,\qquad \bar\delta\leftrightarrow \bar Q,\qquad  
{d\over{dt}}\leftrightarrow H,
\label{ccuatro}
\ee  ($H$ stands for the Hamiltonian operator) implies that:
\be Q^2=\bar Q^2=0,\qquad \{Q,\bar Q\}=Q\bar Q+\bar Q Q=H,
\label{ccinco}
\ee which is the standard supersymmetry algebra for $0+1$-supersymmetric 
field theories. We can carry out explicitly the canonical quantization of
the  theory by imposing the canonical commutation relations:  
\be
\{\bar\psi^\mu,\psi^\nu\}=g^{\mu\nu},\qquad \{\psi^\mu,\psi^\nu\}=
\{\bar\psi^\mu,\bar\psi^\nu\}=0.
\label{cseis}
\ee From these equations it is natural to interpret $\bar\psi$ as fermion
creation operators. In view of this, we have the following structure on 
the Hilbert space:
\be
\cases{ &--States~ with~ one~ fermion:  
$\omega_\mu(x)\bar\psi^\mu|\Omega\rangle$,\cr\cr &--States~ with~ two~
fermions:  
$\omega_{\mu\nu}(x)\bar\psi^\mu\bar\psi^\nu|\Omega\rangle$,
\cr&\,\,\,\,\,\,\,\,\,\,\,\,\,\,\,\,\,\,\,\,\,\,\,\,\,\,\,\,\,\vdots\cr  
&--States~ with~ $n$~ fermions:  
$\omega_{\mu_1,\ldots,\mu_n}(x)\bar\psi^{\mu_1}\cdots
\bar\psi^{\mu_n}|\Omega\rangle$,\cr }
\label{csiete}
\ee being $|\Omega\rangle$ the Clifford vacuum. The Hilbert space of our  
system is thus $\Omega^*(X)$,  the set of differential forms on $X$.
 $Q$ and $\bar Q$ are represented on this Hilbert space by  the exterior
derivative and its adjoint, 
\be Q\leftrightarrow d,\qquad \bar Q\leftrightarrow d^{+},
\label{cocho}
\ee therefore, the Hamiltonian is the Hodge-de Rham Laplacian on $X$:
\be H=dd^{+}+d^{+}d=\Delta.
\label{cnueve}
\ee The zero-energy states are in one-to-one correspondence with the 
harmonic forms on $X$. After  rescaling the parameter $t$ and the
fermionic  fields by, 
\be t\to\gamma^2t,\qquad \bar\psi\to\gamma^{-1}\bar\psi,\qquad  
\psi\to\gamma^{-1}
\psi,
\label{cdiez}
\ee the partition function (\ref{noventasiete}) takes the form:
\be 
\chi_V=\int {{dx d\psi 
d\bar\psi}\over{(2\pi)^{m}}}{\gamma^{2m}\over{\sqrt{g}}}
\ex^{ -\oint_0^{1/\gamma^2} dt \left[\half g_{\mu\nu}\dot x^\mu \dot x^\nu 
-i\bar\psi\nabla_t \psi-{1\over4}
R^{\rho\sigma}{}_{\mu\nu}\bar\psi_\rho\bar\psi_\sigma\psi^\mu
\psi^\nu\right] },
\label{conce}
\ee  Using heat-kernel techniques \cite{luisdos,friedan} one finds:
\be
\chi_V=\tr \left[(-1)^F \ex^{-{1\over{\gamma^2}}H}\right],
\label{cdoce}
\ee where $F$ is the fermion number operator. In the limit $\gamma\to   0$
only the zero-modes of $H$ survive and therefore one must count   harmonic
forms  with signs, which come from $(-1)^F$, leading to the result:
\be
\chi_V=\sum_{k=0}^{2m}(-1)^k b_k =\chi(X),
\label{ctrece}
\ee ($b_k$ are the Betti numbers of $X$) in perfect agreement with our 
previous calculation. 

Actually, due to supersymmetry, for each non-zero energy bosonic mode 
there is  a fermionic one with the same energy which cancels its
contribution to (\ref{cdoce}). Therefore, the computation performed in the
Hamiltonian formalism holds for any $\gamma$.

\subsubsection{Topological sigma models}

Our next example of TQFT was introduced by Witten \cite{tsm} as a  twisted
version of the $N=2$ supersymmetric sigma model in two dimensions.  Here
we will briefly analyze it within the framework of the Mathai-Quillen 
formalism. The model, in its more general form, is defined in terms of a
smooth, almost-hermitian manifold $X$, with metric $G_{mn}$ and 
almost-complex structure $J^i{}_j$ satisfying: 
\be  J^i{}_j J^j{}_k =-\delta^i{}_k,\qquad G_{ij}J^i{}_k J^j{}_m=G_{km}.
\label{ccatorce}
\ee Let us consider the set of smooth maps from a Riemann surface  
$\Sigma$ to $X$, $\phi:\Sigma\to X$, and the vector bundle ${\cale}\to  
\Sigma$ with fibre
${\cal F}=\Gamma\bigl(T^*(\Sigma)\otimes \phi^*(TX)\bigr)^{+}$, where by
$^+$ we denote the self-dual part, \ie, if
$\varrho^i_\alpha\in{\cal F}$, then $\varrho^i_\alpha$ is self-dual, \ie,
$\varrho^i_\alpha J^j{}_i\epsilon_\beta{}^\alpha=\varrho^j_\beta$.
 The choice of section in ${\cale}$ is the following:
\be s(\phi)^i_\alpha = \partial_\alpha x^i +
J^i{}_j\epsilon_\alpha{}^\beta\partial_\beta x^j.
\label{cquince}
\ee Notice that it satisfies the self-duality condition
$s(\phi)^j_\beta = s(\phi)^i_\alpha J^j{}_i \epsilon_\beta{}^\alpha$.

We will restrict the discussion to the simplest case in which the  manifold
$X$ is K\"ahler. Following the general pattern  (\ref{cincuentanueve}), 
the gauge fermion is given by:
\be
\Psi(\phi,\chi,\varrho,B)=\half \intl_\Sigma d^2\sigma\sqrt{h}\left[
\varrho^\alpha_\mu(B^\mu_\alpha +2is^\mu_\alpha  
+\Gamma^\mu_{\nu\sigma}\chi^\nu\varrho^\sigma_\alpha)\right].
\label{cdiezseis}
\ee The model is invariant under the symmetry transformations:
\bea &&\delta x^\mu=\chi^\mu,\qquad  
\delta\varrho^\mu_\alpha=B^\mu_\alpha,\nonumber\\
&&\delta\chi^\mu=0,\qquad\,\,\delta B^\mu _\alpha= 0.
\label{cdiezsiete}
\eea After integrating out the auxiliary fields the action
($A=\delta\Psi$) reads:
\bea &&A(\phi,\chi,\varrho)=\int\limits_{\Sigma} 
d^2\sigma{\sqrt{h}}\,\biggr (\,\half G_{\mu\nu}
h^{\alpha\beta}\partial_\alpha x^\mu\partial_\beta  x^\nu
+\half\epsilon^{\alpha\beta}J_{\mu\nu}\partial_\alpha  x^\mu\partial_\beta
x^\nu
\nonumber\\& &\,\,\,\,\,\,\,\,\,\,\,\,\,\,\,\,\,\,\,\,\,
\,\,\,\,\,\,\,\,\,\,\,\,\,\,\,\,\,\,\,\,\,\,\,\,\,\,\,\,\,\,\,\,\,
-ih^{\alpha\beta}G_{\mu\nu}\varrho^\mu_\alpha \deriv_\beta
\chi^\nu-{1\over8}
h^{\alpha\beta}R_{\mu\nu\sigma\tau}\varrho^\mu_\alpha\varrho^\nu_\beta
\chi^\sigma
\chi^\tau\,\biggl)
\label{cdiezocho}
\eea where $\deriv_\alpha \chi^\mu =\partial_\alpha  
\chi^\mu+\Gamma^\mu_{\nu\sigma}
\partial_\alpha x^\nu\chi^\sigma$.

Rewriting (\ref{cquince}) in terms of holomorphic indices,
$\alpha\to(z,\bar z)$, and $i\to(I,\bar I)$, the equation for the zero
locus of the section  becomes:
\be
\partial_\alpha x^i + J^i{}_j\epsilon_\alpha{}^\beta\partial_\beta
x^j=0\to\partial_{\bar z} x^I=0,
\label{cdieznueve}
\ee
\ie, it corresponds to holomorphic instantons. In order to study  the
dimension of this moduli space one must study  the possible deformations
of the solutions of (\ref{cdieznueve}):
\be x^i\to x^i+\delta x^i,\qquad (\deriv_\alpha \delta x^i)^{+}=0.
\label{cveinte}
\ee This is precisely the field equation for the field  $\chi$, 
\be (\deriv_\alpha \chi)^{+}=0,
\label{cveinteuno}
\ee which clarifies the role played by the $\chi$-zero modes. The
dimension of the moduli space of holomorphic  instantons can be obtained
with the help of an index theorem,  and in many  situations coincides with
the ghost-number anomaly of the theory. Contrary to the case of
supersymmetric quantum mechanics, this anomaly is in general not zero.
This implies that, in general, one is  forced to insert operators to
obtain non-trivial results. 

The observables are obtained from the analysis of  the
$\delta$-cohomology associated to the symmetry (\ref{cdiezsiete}).  The
highest-ghost-number ones turn out to be \cite{tsm}:
\be {\cal O}^{(0)}_A=
A_{i_1,\ldots,i_p}\chi^{i_1}\chi^{i_2}\cdots\chi^{i_p},\qquad  A\in
\Omega^{*}(X),
\label{cveintedos}
\ee and satisfy the relation:
\be
\{Q,{\cal O}^{(0)}_A\}={\cal O}^{(0)}_{dA},
\label{cveintetres}
\ee where $Q$ denotes the generator of the symmetry $\delta$. This 
relation allows to identify the $Q$-cohomology classes of the 
highest-ghost-number observables with the (de Rham) cohomology
 classes of $X$.

The topological descent equations (\ref{vseis}) now take the form:
\be d{\cal O}^{(0)}_{A}=\{Q,{\cal O}^{(1)}_A\},\qquad d{\cal  
O}^{(1)}_{A}=\{Q,{\cal O}^{(2)}_A\}.
\label{cveintecinco}
\ee They are easily solved:
\be {\cal O}^{(1)}_A=A_{i_1,\ldots,i_p}\partial_\alpha  
x^{i_1}\chi^{i_2}\cdots\chi^{i_p} d\sigma^\alpha,\qquad  {\cal
O}^{(2)}_A=\half A_{i_1,\ldots,i_p}\partial_\alpha x^{i_1}\partial_\beta
x^{i_2}\chi^{i_3}\cdots\chi^{i_p}d\sigma^\alpha\wedge d\sigma^\beta.
\label{cveintecuatro}
\ee With the help of these operators one completes the family of 
observables which, as expected, are labeled by homology classes of the 
two-dimensional manifold $\Sigma$:
\be
\intl_\gamma {\cal O}^{(1)}_{A},\qquad \intl_\Sigma {\cal  O}^{(2)}_{A}.
\label{cveinteseis}
\ee 

The topological sigma model which has been described in this section is
called of type A. It turns out that there are two possible ways to twist
$N=2$ supersymmetric sigma models. One of the possibilities leads to
type-A models while the other generates what are called type-B
models \cite{pablo,witmirror}. The existence of these two models is
linked \cite{witmirror} to mirror symmetry in the context of string theory.
The type-B model constitutes a special kind of TQFT which does not fall into
any of the two kinds described in the previous section. Type-A models
depend on the K\"ahler class of the target manifold and are independent of
the complex structure. On the contrary, type-B models depend on the
complex structure and are independent of the K\"ahler class. Type-B models
have been generalized to accommodate Kodaira-Spencer deformation
theory \cite{ks}. They have been also analyzed from other points of
view \cite{yank}.

Topological sigma models have been generalized including potential
terms \cite{vafapo,potenciales}. The resulting theories for the case of
type A have been understood recently in the context of the Mathai-Quillen
formalism after the construction of equivariant extensions \cite{eq}.

\vfill
\newpage

\section{Donaldson-Witten Theory}
\setcounter{equation}{0}

Donaldson-Witten theory was historically the first TQFT to be introduced. 
It was  constructed by Witten \cite{tqft} in 1988 using some insight from
Floer theory \cite{floer} and twisting $N=2$ supersymmetric Yang-Mills
theory. The vacuum expectation values of its observables are Donaldson
invariants for four-manifolds \cite{donaldcero,donald,donaldbook}. The
theory was later analyzed by Atiyah and  Jeffrey from the viewpoint of  the
Mathai-Quillen formalism \cite{jeffrey}. 
See  \cite{lape} for other early approaches to this theory.

Donaldson invariants were introduced \cite{donaldcero} by S. Donaldson in
1983. They are topological invariants for four-manifolds which depend on
the differentiable structure of the manifold. They are very important in
topology because they are helpful in the classification of differentiable
four-manifolds. Contrary to the case of dimensions two and three, for
higher dimensions there are topological obstructions for the existence of
smooth structures. Though the origin of this problem is well understood in
dimensions five and higher, the situation in four dimensions is quite
different. Donaldson invariants constitute a very promising tool to
improve our knowledge in the case of four dimensions.

The geometric framework for Donaldson-Witten theory is the following. Let
us consider  a compact oriented four-dimensional manifold $X$ endowed with
a  metric $g_{\mu\nu}$. Over this manifold $X$ we construct a principal
bundle,
$P\to X$, with group $G$ which will be assumed to be simple and compact.  
The  automorphism group of the bundle $P$, ${\cal G}$, is the infinite  
dimensional gauge group, whose Lie algebra will be denoted by
${\hbox{\rm  Lie}}({\cal G})=\Gamma({\hbox{\rm ad}}P)=\Gamma(P\times_{\rm
ad}{\bf g})=\Omega^0(X,\ad P)$. A connection in $P$ will be denoted by $A$
and the corresponding covariant derivative  and self-dual part of its
curvature by
$D_\mu$ and $F^+=p^{+}(dA+A\wedge A)$, respectively. 

The aim of Donaldson-Witten theory is to reformulate, in a  field-theoretic
language, the theory proposed by Donaldson \cite{donaldcero,donald}  which
characterizes diffeomorphism classes of four-manifolds  in terms of
cohomology classes built on the moduli space of the anti-self-dual (ASD)
$G$-instantons of Atiyah, Hitchin and  Singer \cite{ahs,freeddos}. The main
ingredient of the theory is therefore the instanton equation:
\be  F^{+}(A)={1\over2}\left(\,F(A)+*F(A)\,\right)=0.
\label{menelao}
\ee  Starting from the instanton equation (\ref{menelao}) we would like to 
build the topological field theory which is associated to these equations
in the framework of the Mathai-Quillen formalism. According to our previous
discussion,  we have to specify (i) a  field space, and (ii) a vector
space with the quantum numbers of equation  (\ref{menelao}). The
configuration or field space of the theory is just the space of
$G$-connections on $P$, ${\cal A}$. The vector space
${\cal F}$ is the space of self-dual two-forms on $X$ with values in the
adjoint bundle $\ad P$, $\Omega^{2,+}(X,\ad P)$. There is a natural 
action of the group of gauge transformations ${\cal G}$ on both
${\cal A}$ and ${\cal  F}$, which allows us to introduce the principal
bundle ${\cal A}\to {\cal A}/{\cal  G}$, and the associated vector bundle
${\cale}_+= {\cal A}\times_{\cal G}\Omega^{2,+}(X,\ad P)$. In this context,
equation (\ref{menelao}) is regarded as defining a section of
${\cale}_+$,  $s:{\cal A}/{\cal G}\to {\cale}_+$,
\be s(A)=F^+(A).
\label{seccion}
\ee The zero locus of this section gives precisely the moduli space of ASD
instantons. 

In order to complete the construction we must  specify the field content
of the theory. Let us introduce the following set of fields:
\begin{equation}
\chi_{\mu\nu}, \; G_{\mu\nu} \; \in \; \Omega^{2,+}(X,\ad P),
\;\;\;\;\;
\psi_\mu \; \in \; \Omega^1(X,\ad P),
\;\;\;\;\;
\eta, \; \lambda, \; \phi \; \in \; \Omega^{0}(X,\ad P).
\label{cveintesiete}
\end{equation} The ghost number  carried by each of the fields is the
following:

\vspace{1pc}
\begin{center}
\begin{tabular}{||c|c|c|c|c|c|c||}
\hline
$A_\mu$ & $\chi_{\mu\nu}$ & $G_{\mu\nu}$ & $\psi_\mu$ & $\eta$ &  $  
\lambda$ &
$\phi$\\
\hline
$0$ & $-1$ & $0$ & $1$ & $-1$ & $-2$ & $2$ \\
\hline
\end{tabular}
\end{center}
\vspace{1pc}

\noindent This field content is bigger than the standard one described in
the previous section in our discussion of supersymmetric quantum mechanics
and topological sigma models. The reason for this is that for situations
in which a  gauge symmetry is present the Mathai-Quillen formalism must be
modified so  that pure gauge degrees of freedom are projected out. We will
not discuss these  aspects here. We refer the reader to  \cite{moore} for
details. The outcome of  the analysis is that now the gauge fermion
decomposes into two parts:
$\Psi=\Psi_{\rm loc}+\Psi_{\rm proj}$. The first one enforces  the
localization into the moduli space while the second one takes care of  the
projection.

In (\ref{cveintesiete}) the Grassmann-odd self-dual two-form  
$\chi_{\mu\nu}$ is the fibre antighost, while 
$G_{\mu\nu}$ is its bosonic partner (it is an auxiliary field). The  
Grassmann-odd one-form
$\psi_\mu$  lives in the (co)tangent space to the field space and is to  be
understood as providing a basis for differential forms on ${\cal A}$, 
whereas the scalar bosonic field
$\phi$ --or rather its expectation value
$\langle
\phi\rangle$--  plays the role of the curvature two-form of the bundle  
${\cal A}\to{\cal A}/{\cal G}$. The Grassmann-odd scalar field, $\eta$,  
together with its bosonic partner,
$\lambda$, enforce the horizontal projection \cite{moore}. 

The scalar symmetry which characterizes the theory has the form:
\begin{eqnarray}
\delta A_\mu &= \psi_\mu,  \mbox{\hskip2cm} \delta \chi_{\mu\nu} &=  
G_{\mu\nu},
\nonumber \\ \delta \psi &= d_A\phi, \mbox{\hskip1.5cm}
\delta G_{\mu\nu} &= i[\chi_{\mu\nu},\phi], \nonumber \\
\delta \phi &= 0, \mbox{\hskip2.3cm}
\delta\lambda &= \eta,
\mbox{\hskip2cm} \delta\eta = i [\lambda,\phi],
\label{cveintenueve}
\end{eqnarray} where $\delta^2=$ gauge transformation with gauge parameter
$\phi$, so  one is led to study the  ${\cal G}$-equivariant cohomology of
$\delta$. The  action of the theory is $\delta$-exact.  The appropriate
gauge fermions are:
\bea
\Psi_{\rm loc}&=&\intl_X d^4 x \sqrt{g} \tr\left[2\chi_{\mu\nu}(
F^{+\mu\nu}-\half G^{\mu\nu})\right],\nonumber\\
\Psi_{\rm  proj}&=&\intl_X d^4 x \sqrt{g} \tr\left[i\lambda  D_\mu\psi^\mu
\right].
\label{cveinteocho}
\eea After integrating out the auxiliary fields the action reads:
\bea
\delta(\Psi_{\rm loc}+\Psi_{\rm  proj}) &\to& \int_M d^4
x\,\sqrt{g}\tr\Big({F^+}^2-i\chi^{\mu\nu} D_\mu \psi_\nu +i\eta D_\mu
\psi^\mu +\frac{1}{4} \phi \{ \chi_{\mu\nu},\chi^{\mu\nu} \} \nonumber  
\\ &&\,\,\,\,\,\,\,\,\,\,\,\,\,\,\,\,\,\,\,\,\,\,\,\,\,\,\,\,\,\,\,\,\,\,\,
\,\,\,\,\,\,\,\,\,\,\,\,\, +\frac{i}{4} \lambda \{\psi_\mu,\psi^\mu\}
-\lambda D_\mu D^\mu\phi \Big).
\label{ctreinta}
\eea

The moduli space associated to the theory is the space of solutions of
(\ref{menelao}) modulo gauge transformations. This is the space of  ASD
instantons, which is finite dimensional and will be denoted by  
${\cal M}_{\rm ASD}$. To obtain its dimension one has to study the number
of independent perturbations to the equations (\ref{menelao}), modulo
gauge transformations. They are given by the equations:
\bea A\to A+\delta A \Rightarrow (D_\mu \delta A_\nu)^+&=&0,\nonumber\\
d^{*}_A
\delta A&=&0.
\label{ctreintados}
\eea The second equation just says that $\delta A$ is orthogonal to the  
vertical directions (gauge orbits) tangent to the field space, which are
of the  form $d_{A}
\omega$, $\omega\in \Omega^0(X,\ad P)$. The equations above are precisely 
the
$\psi_\mu$-field equations as derived from the action (\ref{ctreinta}):  
$(D_\mu\psi_\nu)^+=0$, $D_\mu\psi^\mu=0$, ($\psi_\mu$ zero modes). The  
dimension of the moduli space is calculated from (\ref{ctreintados}) with
the aid of an  index theorem \cite{ahs}. For $G=SU(2)$, the result is:
\be  {\rm dim}\,{\cal M}_{\rm ASD}=8k-{3\over2}(\chi +\sigma),
\label{indice}
\ee where $k$ is the instanton number, and $\chi$ and $\sigma$ are,  
respectively, the Euler characteristic and the signature of the manifold
$X$. The  ghost-number anomaly equals precisely the dimension of the moduli
space, which is  generically not zero. This implies that observables must
be introduced to obtain non-vanishing vacuum expectation values.  The
observables of the theory are obtained from the analysis of the  
${\cal G}$-equivariant cohomology of
$\delta$ (recall $\delta^2=$ gauge transformation). We will come back  to
this issue later. Now we will construct the theory from a different point 
of view.

\subsection{Twist of $N=2$ supersymmetry}

We have described Donaldson-Witten theory from the point of view of the
Mathai-Quillen formalism. The construction results rather compact and
geometric. However, this approach was not available in the  early  days of
TQFT, and, in fact, the theory was originally constructed in the less
geometric way that we will review now. This alternative formulation,
though being less transparent from the geometric point  of view, provides
an explicit link to four-dimensional
$N=2$ supersymmetric  Yang-Mills theory which has proved to be very
fruitful to perform explicit calculations. 

Let us begin with a review of some generalities  concerning $N=2$
supersymmetry in  four-dimensions. The global symmetry group of $N=2$
supersymmetry in
${\RR}^4$ is 
$H= SU(2)_L\otimes SU(2)_R \otimes SU(2)_I \otimes U(1)_{\cal R}$  where
${\cal K} = SU(2)_L \otimes SU(2)_R$ is the rotation group and 
$SU(2)_I \otimes U(1)_{\cal R}$ is the internal (chiral) symmetry  group.
The supercharges, $Q^i_\alpha$ and $\overline Q_{i\dot\alpha }$, which 
generate
$N=2$ supersymmetry have the following transformations under $H$:
\begin{equation} Q^i_\alpha \;\; (\frac{1}{2},0,\frac{1}{2})^1,
\;\;\;\;\;\;\;\;\;\;\;
\overline Q_{i \dot\alpha } \;\;
 (0,\frac{1}{2},\frac{1}{2})^{-1},
\label{ctreintatres}
\end{equation} where the superindex denotes the $U(1)_{\cal R}$ charge and
the numbers within parentheses the representations under each of the
factors in 
$SU(2)_L\otimes SU(2)_R \otimes SU(2)_I$. The supercharges  
(\ref{ctreintatres}) satisfy: 
\be
\{Q^i_\alpha, \overline Q_{j\dot\beta} \} = \delta^i_j 
P_{\alpha\dot\beta}.
\label{ctreintacuatro}
\ee

The twist consists of considering as the rotation group the group,
${\cal K}' = SU(2)_L'\otimes SU(2)_R$, where $SU(2)_L'$ is the diagonal  
subgroup of $SU(2)_L\otimes SU(2)_I$. This implies that the isospin index
$i$ becomes a spinorial index $\alpha$: $Q^i_\alpha \rightarrow
Q^\beta_\alpha$ and
$\overline Q_{i\dot\beta } \rightarrow G_{\alpha\dot\beta}$.  Precisely the
trace of $Q^\beta_\alpha$ is chosen as the generator of the scalar 
symmetry:
$Q = Q^\alpha_\alpha$. Under the new global group $H'={\cal K}'\otimes
U(1)_{\cal R}$, the symmetry generators transform as:
\begin{equation} G_{\alpha\dot\beta} \;\; (\frac{1}{2},\frac{1}{2})^{-1},
\;\;\;\;\;\;\;\;\;\; Q_{(\alpha\beta)} \;\; (1,0)^1,
\;\;\;\;\;\;\;\;\;\; Q \;\; (0,0)^1.
\label{ctreintacinco}
\end{equation}

Notice that we have obtained a scalar generator $Q$. It is important to 
stress that as long as we stay on a flat space (or one with trivial
holonomy),  the twist is just a fancy way of considering the theory, for
in the end we  are not changing anything. However, the appearance of a
scalar symmetry  makes the procedure meaningful when we move to an
arbitrary four-manifold.  Once the scalar symmetry is found we must study
if, as stated in (\ref{angela}), the energy-momentum tensor is exact, \ie,
if it can be written as the transformation of some quantity under $Q$. The
$N=2$ supersymmetry algebra gives a  necessary condition for this to 
hold. Indeed, after the twisting, this algebra becomes:
\begin{equation}
\{Q^i_\alpha, \overline Q_{j\dot\beta} \} = \delta^i_j  P_{\alpha\dot\beta}
\longrightarrow
\{ Q , G_{\alpha\beta} \} = P_{\alpha\dot\beta},\qquad \{Q,Q\}=0,
\label{ctreintaseis}
\end{equation} where $P_{\alpha\dot\beta}$ is the momentum operator of the
theory. Certainly (\ref{ctreintaseis}) is only a necessary condition for
the theory to be  topological. However, up to date, for all the $N=2$
supersymmetric  models whose twisting has been studied the relation on the
right hand side of (\ref{ctreintaseis}) has become valid for the whole
energy-momentum tensor. Notice that (\ref{ctreintaseis}) are the basic
equations (\ref{once}) and (\ref{trece}) of our general discussion on TQFT.
It is important to remark that twisted theories are considered as
Euclidean theories. This implies that the twisting procedure is often
accompanied by some changes on the complex nature of the fields. This
delicate issue has been treated recently by Blau and
Thompson \cite{euclides}.

In ${\RR}^4$ the original and the twisted theories are equivalent. 
However, for arbitrary manifolds they are certainly different due to the
fact that  their energy-momentum tensors are not the same. The twisting
changes the spin  of the fields in the theory, and therefore their
couplings  to the metric on
$X$ become modified. This suggests an alternative way of  looking at the
twist. All that has to be done is: gauge the internal group
$SU(2)_I$, and identify the corresponding $SU(2)$ connection with the spin
connection on $X$. This process changes the spin connection and therefore
the energy-momentum tensor of the theory,  which in turn modifies the
couplings  to gravity of the different fields of the theory. This
alternative point of view to the twisting procedure has been recently
reviewed in this context in
 \cite{baryon}. 

Under the twist, the field content is modified as follows:
\bea A_{\alpha\dalpha}\,\,(\half,\half,0)^0 &\too&
A_{\alpha\dalpha}\,\,(\half,\half)^0,\nonumber\\
\lambda_{\alpha i}\,\,(\half,0,\half)^{-1} &\too&
\chi_{\alpha\beta}\,\,(1,0)^{-1},\,\,\eta (0,0)^{-1},\nonumber\\
\bar\lambda^j_{\dalpha}\,\,(0,\half,\half)^1 &\too&
\psi_{\alpha\dalpha}\,\,(\half,\half)^1,\\  B\,\,\,(0,0,0)^{-2} &\too&
\lambda\,\,\,(0,0)^{-2},
\nonumber\\ B^{*}\,\,\,(0,0,0)^2 &\too&
\phi\,\,\,(0,0)^2,
\nonumber\\ D_{ij}\,\,(0,0,1)^0 &\too& G_{\alpha\beta}\,\,(1,0)^0.\nonumber
\label{ctreintasiete}
\eea In the process of twisting, the $U(1)_{\cal R}$ symmetry becomes the  
$U(1)$-like symmetry associated to the ghost number of the topological
theory. The  ghost number anomaly is thus naturally related to the chiral
anomaly of
$U(1)_{\cal R}$. The twisted action differs from the action
(\ref{ctreinta}) obtained in the Mathai-Quillen formalism by a term of the
form,
\be
\int_M d^4 x\,\sqrt{g}\tr\Big( \frac{i}{2}\phi\{\eta,\eta\}
+\frac{1}{8}[\lambda,\phi]^2 \Big).
\label{rourke}
\ee This term turns out to be $Q$-exact ($\sim
\{Q,\int\eta[\phi,\lambda]\}$) and therefore it can be ignored.

Associated to each of the independent Casimirs of the gauge group $G$  it
is possible to construct highest-ghost-number operators. For example, for 
the quadratic Casimir this operator is:
 \be W_0 = \frac{1}{8\pi^2} \tr (\phi^2), 
\label{raquela}
\ee  and it generates the following family of operators:
\be W_1 = \frac{1}{4\pi^2}\tr(\phi \psi), \,\,\,\,\,\,\,\,\,\,\,\,\,\, W_2
=
\frac{1}{4\pi^2} \tr(\frac{1}{2}\psi\wedge\psi+\phi\wedge F),
 \,\,\,\,\,\,\,\,\,\,\,\,\,\, W_3 = \frac{1}{4\pi^2} \tr(\psi\wedge F).
\label{raquel}
\ee These operators are easily obtained by solving the descent equations,  
$\delta W_i = d W_{i-1}$. From them one defines the following observables:
\begin{equation} {\cal O}^{(k)} = \int_{\gamma_k} W_k,
\label{sonia}
\end{equation} where $\gamma_k \in H_k(M)$. The descent equations imply
that they are
$\delta$-invariant and that they only depend on the homology class
$\gamma_k$.

The functional integral corresponding to the topological invariants  of 
the theory has the form:
\begin{equation}
\langle {\cal O}^{(k_1)} {\cal O}^{(k_2)} \cdots {\cal O}^{(k_p)} \rangle =
\int  {\cal O}^{(k_1)} {\cal O}^{(k_2)} \cdots {\cal O}^{(k_p)} \exp
({-{S}/{g^2}}),
\label{clara}
\end{equation} where the integration has to be understood on the space of
field  configurations modulo gauge transformations and $g$ is a coupling
constant. The standard arguments described in sect. 2 show that due to the $\delta$-exactness of
the action $S$, the  quantities obtained in (\ref{clara}) are independent
of $g$. This implies that  the observables of the theory can be obtained
either in the limit
$g\rightarrow 0$, where perturbative methods apply, or in the limit
$g\rightarrow \infty$, where one is forced to consider a  non-perturbative
approach. The crucial point is to observe that the $\delta$-exactness of
the  action implies, at least formally, that in either case the values of
the vevs must be the  same. 

\subsection{Perturbative approach}

The previous argument for $g\rightarrow 0$ implies that the  semiclassical
approximation of the theory is exact. In this limit the contributions  to
the functional integral are dominated by the bosonic field configurations 
which minimize
$S$. These turn out to be given by the equations:
\be  F^+=0,\qquad D_\mu D^\mu\phi=0.
\label{ctreintaocho}
\ee Let us assume that in the situation under consideration there are only
irreducible connections  (this is true in the case $b^+_2={\rm
dim}\,\Omega^{2,+}(X)>1$). In this case  the contributions from the even 
part of the action are given entirely by  the solutions of the equation
$F^+=0$,
\ie, by instanton configurations. Being the connection irreducible  there
are no non-trivial  solutions to the second equation in
(\ref{ctreintaocho}).

The zero modes of the field $\psi$ come from the  solutions to the
equations,
\begin{equation} (D_\mu\psi_\nu)^+=0, \;\;\;\;\;\;\;\; D_\mu\psi^\mu=0,
\label{ctreintanueve}
\end{equation} which are precisely the ones that define the tangent space
to the space of instanton  configurations. The number of independent
solutions of these equations determine the  dimension of the instanton
moduli space ${\cal M}_{\rm ASD}$. As stated in  (\ref{indice}), for
$SU(2)$,  $d_{{\cal M}_{\rm ASD}} = 8k-3(\chi+\sigma)/2$. 

The fundamental contribution to the functional integral (\ref{clara})  is
given by the elements of ${\cal M}_{\rm ASD}$ and by the zero-modes of the
solutions to (\ref{ctreintanueve}). Once these have been obtained they
must be introduced in the action and an expansion up to quadratic terms in
non-zero modes  must be performed. The fields $\phi$ and $\lambda$ are
integrated out originating a contribution \cite{tqft} which is equivalent
to the replacement of the field $\phi$ in the operators ${\cal O}^{(k)}$
by,
\be
\langle\phi^a\rangle=\int d^4y
\sqrt{g}G^{ab}(x,y)[\psi_\mu(y),\psi^\mu(y)]^b,
\label{smock}
\ee where $G^{ab}(x,y)$ is the inverse of the Laplace operator,
\be D_\mu D^\mu  G^{ab}(x,y)=\delta^{ab}\delta^{(4)}(x-y).
\label{propa}
\ee These are the only relevant terms in the limit
$g\rightarrow  0$. The resulting gaussian integrations then must be
performed. Due to the  presence of the $\delta$ symmetry these come in
quotients whose value is $\pm 1$. The functional integral (\ref{clara})
takes the form: 
\begin{eqnarray} && \langle {\cal O}^{(k_1)} {\cal O}^{(k_2)} \cdots {\cal
O}^{(k_p)} \rangle = \nonumber \\ && \int_{{\cal M}_{\rm ASD}} da_1\cdots
da_{d_{{\cal M}_{\rm ASD}}} d\psi_1
\cdots d\psi_{d_{{\cal  M}_{\rm ASD}}} {\cal O}^{(k_1)} {\cal O}^{(k_2)}
\cdots {\cal O}^{(k_p)}  (-1)^{\nu(a_1,\dots,a_{d_{{\cal M}_{\rm
ASD}}})},\nonumber \\
\label{clarados}
\end{eqnarray} where $\nu(a_1,\dots,a_{d_{{\cal M}_{\rm ASD}}})=0,1$. The
integration over the  odd modes leads to a selection rule for the product
of observables. This selection rule  is better expressed making use of the
ghost numbers of the fields. For the operators  in (\ref{sonia}) one
finds: $U({\cal O}^{(k)})=4-k$, and  the  selection rule can be written as
$d_{{\cal M}_{\rm ASD}}=\sum_{i=1}^p U({\cal O}^{(k_i)})$.

In the case in which $d_{{\cal M}_{\rm ASD}}=0$, the only observable is
the partition function, which takes the form:
\begin{equation}
\langle 1 \rangle = \sum_i (-1)^{\nu_i},
\label{vanessa}
\end{equation} where the sum is over isolated instantons, and $\nu_i=\pm
1$. In  general, the integration of the zero-modes in  (\ref{clarados})
leads to an antisymmetrization in such a way that one ends with the
integration of  a
$d_{{\cal M}_{\rm ASD}}$-form on ${\cal M}_{\rm ASD}$. The resulting real
number is a  topological invariant. Notice that in the process a map 
\be H_k(M) \longrightarrow  H^k({\cal M}_{\rm ASD})
\label{prodi}
\ee has been constructed. The vevs of the theory provide polynomials in
$H_{k_1}(M) \times H_{k_2}(M) \times \cdots \times  H_{k_p}(M)$ which are
precisely the Donaldson polynomials invariants.

\subsection{Non-perturbative approach}

The study of Donaldson-Witten theory from a perturbative point of view 
proved that the vevs of the observables of this theory are related to
Donaldson invariants. However, it did not  provide a new method to compute
these invariants since the functional integral  leads to an integration
over the moduli space of instantons, which is precisely the  step where
the hardest problems to compute these invariants appear. In the context of
quantum field theory there exists the possibility of studying  the form of
these observables from a non-perturbative point of view, \ie, in the 
strong coupling limit
$g\rightarrow \infty$. This line of research seemed difficult to  implement
until very recently. However, in  1994, after the work by Seiberg and
Witten \cite{sw}, important progress was made in the knowledge of the
non-perturbative structure of $N=2$ supersymmetric Yang-Mills theories.
Their results were immediately applied to the twisted theory leading to
explicit expressions for the topological invariants in a variety of
situations \cite{abm}. But  perhaps the most important outcome of this
approach is the emergence of the  existence of a relation between the
moduli space of instantons and other moduli  spaces such  as the moduli
space of abelian monopoles which will be introduced  below.

$N=2$ supersymmetric Yang-Mills theory is asymptotically free. This  means
that the effective coupling constant becomes small at large energies. The
perturbative methods which have been used are therefore valid at these
energies.  At low energies, however,  those methods are not valid and  one
must use non-perturbative techniques. Before $1994$ the infrared behavior 
of the
$N=2$ supersymmetric theory was not known and the non-perturbative  
approach seemed to be out of reach. However, the infrared behavior of $N=1$
supersymmetric Yang-Mills was known, and, in 1993, Witten \cite{wijmp} was
able to  make explicit calculations for the Donaldson invariants on
K\"ahler  manifolds with
$H^{(2,0)}\not =0$ using information concerning these theories. This 
approach is known as the abstract approach while the one based on the
infrared behavior of $N=2$ supersymmetric Yang-Mills is referred to as the
concrete approach. We will discuss them now in turn. 

\vspace{1pc}

\subsubsection{Donaldson invariants: abstract approach}

The key ingredient of this approach is the following observation due  to
Witten \cite{wijmp}: on a K\"ahler manifold with $H^{(2,0)}\not =0$
Donaldson-Witten theory can be perturbed by a mass  term preserving the
topological character of the theory. The theory can then  be regarded as a
twisted $N=1$ super Yang-Mills theory with matter fields. The infrared 
behavior of this theory is known: it has a mass gap and undergoes
confinement  and chiral symmetry breaking. Moreover, the ${\ZZ}_{2h}$
subgroup ($h$ is  the dual Coxeter number of the gauge group $G$) of
$U(1)_{\cal R}$  which is preserved by instantons is believed to be  
spontaneously broken to
${\ZZ}_2$, which allows fermion masses, giving rise to an $h$-fold
degeneracy of the vacuum.  Vacuum expectation values are written as a sum
over contributions from  each of the  vacua, these contributions being
related by the broken symmetry ${\ZZ}_h={\ZZ}_{2h}/ {\ZZ}_2$. Witten
studied the case of
$SU(2)$ and  he proved that the vevs have the structure first found by 
Kronheimer and Mrowka \cite{km}. We shall now briefly review the
fundamentals of his  construction.

On a four-dimensional K\"ahler manifold the holonomy is reduced  according
to the pattern:
\be SU(2)_L\otimes SU(2)_R\too U(1)_L\otimes SU(2)_R.
\label{ccuarenta}
\ee The ${\bf 2}$ of $SU(2)_L$ decomposes as a sum of one-dimensional  
representations of $U(1)_L$. In particular, for the $N=2$ supersymmetric
charges  
$Q^i_\alpha$ we have:
\be  Q^i_\alpha\too Q^i_1\oplus Q^i_2.
\label{ccuarentauno}
\ee After twisting these give rise to two independent scalar charges, each
transforming under definite $U(1)_L$ transformations:
\be Q_1=Q_1^1, \qquad\qquad Q_2= Q_2^2.
\label{ccuarentados}
\ee These charges satisfy the relations:
\be  Q=Q_1+Q_2,\qquad (Q_1)^2=0=(Q_2)^2,\qquad \{Q_1,Q_2\}=0.
\label{ccuarentatres}
\ee It is important to remark that from the point of view of the untwisted
theory
$Q_1$ can be regarded in the context of $N=1$ superspace as a  derivative
with respect to $\theta_1$. 

The field content of $N=2$ super Yang-Mills theory consists of a gauge (or
vector) multiplet, which is represented by a constrained chiral spinor
superfield $W_\alpha(A, \lambda^1)$, and a  scalar multiplet, which is
represented by a chiral $N=1$  superfield $\Psi(B,\lambda^2)$. The action
in
$N=1$ superspace takes the form:
\be S=\int d^4 xd^2\theta d^2\bar\theta (\Psi^{\dag}\ex^V \Psi)+
\int d^4 xd^2\theta \tr(W^\alpha W_\alpha)+
\int d^4 xd^2\bar\theta \tr(\bar W_\dalpha \bar W^\dalpha).
\label{ccuarentacuatro}
\ee In this expression $V$ is the vector superpotential, related to $W$
by  
$W\sim
\bar D^2
\ex^{-V}D\ex^V$, and $D$ and $\bar D$ are superspace covariant 
derivatives.  It is well known that
$\tr W^2\big|_{\theta^2}$ and $\tr\bar W^2\big|_{\bar\theta^2}$  coincide
up to a $\theta$-term. This implies that the action
(\ref{ccuarentacuatro}) is
$Q_1$-exact modulo a shift in the $\theta$-angle (since $Q_1$ can be
regarded as a derivative with respect to $\theta_1$). This shift can be
absorbed in a chiral rotation which implies a rescaling of the
observables. 

When $X$ is simply connected, and for gauge group $SU(2)$, the only 
relevant observables in Donaldson-Witten theory are the ones associated to
even forms in (\ref{raquela}) and (\ref{raquel}):
\bea  {\cal O}&=&\frac{1}{8\pi^2}\tr(\phi^2),\nonumber\\ 
I(\Sigma)&=&\frac{1}{4\pi^2}\intl_\Sigma \tr(\phi  F+\half\psi\wedge\psi),
\label{ccuarentacinco}
\eea where $\Sigma$ is a two-cycle on the manifold $X$. Both are
$Q_1$-invariant.  Following Witten \cite{wijmp} we perturb the theory
introducing a mass term for the $\Psi$ superfield:
\be
\Delta S=-\int \omega\wedge d^2\bar z  d^2\theta\tr(\Psi^2)\,+\,{\hbox{\rm
h.c.}},\qquad
\omega\in H^{(2,0)}(X),
\label{ccuarentaseis}
\ee which breaks the symmetry from $N=2$ down to $N=1$. This mass term is
not 
$Q_1$-exact. However, it turns out that the perturbed action has the
following form:
\be S+\Delta S= S+I(\tilde\omega)+\{Q_1,\ldots\}
\label{ccuarentasiete}
\ee being $\tilde\omega$ the Poincare dual to $\omega$. Since  
$I(\tilde\omega)$ is after all an observable, the perturbation only
introduces a relabeling  of the observables themselves. To see this, 
consider the generating function for the Donaldson polynomials:
\be
\langle \ex^{\sum_a \alpha_a I(\Sigma_a)+\lambda{\cal O}}\rangle,
\label{ccuarentaocho}
\ee where $\{\Sigma_a\}_{a=1,\ldots,b_2(X)}$ is a basis of $H_2(X)$, and 
$\alpha_a$ and $\lambda$ are constant parameters. The perturbation
(\ref{ccuarentasiete}) just amounts to a shift in the $\alpha_a$
parameters. 

Summarizing, we have shown that for K\"ahler manifolds with  
$H^{(2,0)}\not =0$ there exists a TQFT, which can be regarded as a twisted
version of $N=1$ supersymmetric Yang-Mills theory, whose vev
(\ref{ccuarentaocho}) differ from the corresponding ones in
Donaldson-Witten theory by a shift in the  parameters $\alpha_a$. We will
now use the knowledge on the infrared behavior of $N=1$ supersymmetric
Yang-Mills theory to compute (\ref{ccuarentaocho}) in the new topological
theory.

The first step consists of a rescaling of the metric,
$g_{\mu\nu}\too tg_{\mu\nu}$. In the limit $t\to\infty$ one expects  that a
description in terms of the degrees of freedom of the vacuum states of  the
physical theory in ${\RR}^4$ is valid. Hence, the idea is to compute  the
observables of the twisted theory on each vacuum of the $N=1$
supersymmetric gauge theory. 

If $\omega$ does not vanish, the untwisted theory possesses $h$ vacuum 
states. Take $G=SU(2)$, $h=2$. Standard arguments based on general
properties  of TQFTs and $N=1$ supersymmetry lead to the following result
for the  generating functional \cite{wijmp}:
\bea
\langle \ex^{\sum_a \alpha_a I(\Sigma_a)+\lambda{\cal O}}\rangle  
&=&C_1\ex^{\left(
{\eta_1\over2}\sum_{a,b}\alpha_a\alpha_b\#(\Sigma_a\cap\Sigma_b)+
\lambda\xi_1\right)}\nonumber\\ &&+C_2\ex^{\left(
{\eta_2\over2}\sum_{a,b}\alpha_a\alpha_b\#(\Sigma_a\cap\Sigma_b)+
\lambda\xi_2\right)},
\label{ccuarentanueve}
\eea (notice that each term in (\ref{ccuarentanueve}) comes from each  of
the two vacua) where,
\be  C_i=\ex^{(a_i\chi +b_i\sigma)},
\label{ccincuenta}
\ee is the partition function of the theory in the $i$ vacuum, and $a_i$,  
$b_i$,
$\eta_i$ and $\xi_i=\langle {\cal O}\rangle_i$ ($i=1,2$) are universal  
constants independent of the manifold
$X$. The symmetry ${\ZZ}_2$ gives relations among the  constants:
\be C_2=i^\Delta C_1,\qquad \eta_2=-\eta_1,\qquad \xi_2=-\xi_1,
\label{ccincuentauno}
\ee with $\Delta=\half d_{\cal M}=4k-{3\over4}(\chi+\sigma)$. The relation
between $\eta_1$ and $\eta_2$, and $\xi_1$ and $\xi_2$,  results very
simply from the ${\ZZ}_2$ transformations of the observables. The  relation
between $C_1$ and $C_2$ is the result of taking into consideration the
gravitational anomaly associated to that symmetry. Notice that  
$i^\Delta$ is independent of the instanton number $k$.

If $\omega$ vanishes along some regions, each vacuum is further split  up
into two along each region \cite{wijmp}. We will not discuss this more
general  case in these lectures. The result agrees with the general
structure found by Kronheimer and Mrowka \cite{km}. The unknown parameters
in (\ref{ccincuenta}), (\ref{ccincuentauno}) are universal,
\ie, independent of $X$, and they can be fixed by comparison to the known
values of (\ref{ccuarentaocho}) for some manifolds. The success of this
approach  has an outstanding importance,
 for the agreement found between the results of the calculation and
previously known mathematical results gives support to  the conjectured
picture in the physical $N=1$ supersymmetric gauge theory.

\vspace{1pc}

\subsubsection{Donaldson invariants: concrete approach}

As explained at the beginning of this subsection, in $1994$,  Seiberg and 
Witten, using arguments based on duality,  obtained exact results \cite{sw} for many $N=2$  supersymmetric
Yang-Mills theories, determining their moduli space of vacua. These
physical results were  immediately applied \cite{abm} to the corresponding
twisted theories obtaining a  new expression for Donaldson invariants in
terms of a new set of  invariants: the Seiberg-Witten invariants.

\begin{figure}                

\centerline{\hskip.4in \epsffile{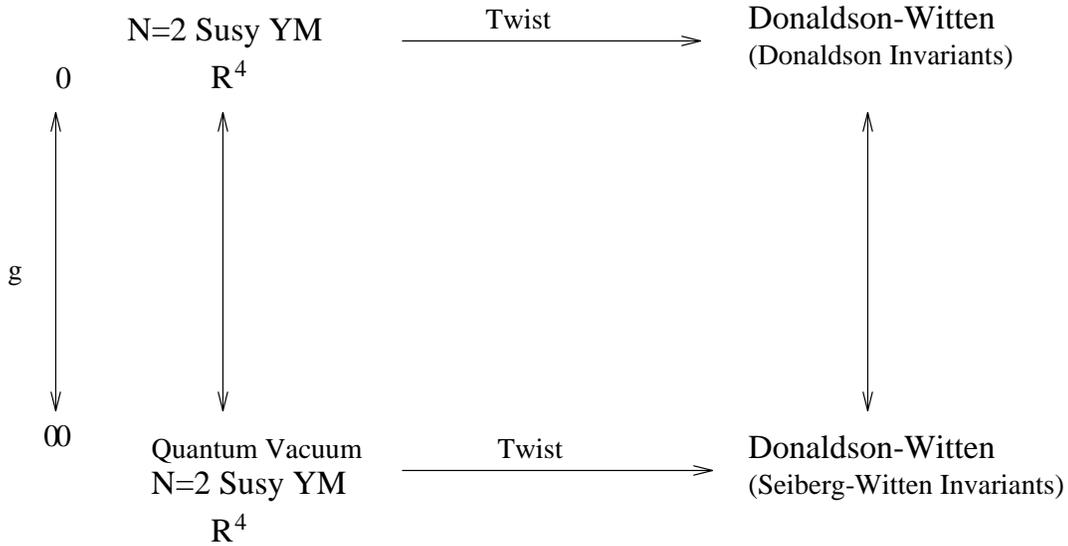}}       

\caption{Schematic diagram of the concrete approach.}            
\label{dw.eps}              

\end{figure}

The general argument which explains why the exact results on the  infrared
behavior of $N=2$ supersymmetric Yang-Mills theories can be used in the  
context of TQFT is the following. In the twisted theory the presence of the
coupling $g$ can be regarded  as a rescaling of the metric. In the limit
$g\to\infty$ the rescaling of the  metric is arbitrarily large and one
expects that calculations can be done in  terms of the vacua corresponding
to
${\RR}^4$.  Recall that $N=2$ supersymmetric Yang-Mills theories are
asymptotically  free and their large distance behavior is equivalent to
their low energy one.  This argument is summarized in Fig.~\ref{dw.eps}.

 From the work by Seiberg and Witten \cite{sw} follows that at low  energies
$N=2$ supersymmetric Yang-Mills theories behave as abelian gauge 
theories. For the case of gauge group $SU(2)$, which will be the case 
considered in this discussion, the effective low energy theory is
parametrized  by a complex variable $u$ which labels the vacuum structure
of the theory.  At each value of
$u$ the effective theory is an $N=2$ supersymmetric abelian gauge theory
coupled to $N=2$ supersymmetric matter fields. One of the most salient
features of the effective theory is that there are points in the
$u$-complex plane where some matter fields become massless. These  points
are singular points of the vacuum moduli space and they are located at
$u=\pm\Lambda^2$, where $\Lambda$ is the dynamically generated scale of the
theory.  At $u=\Lambda^2$ the effective theory consists of an  
$N=2$ supersymmetric abelian gauge theory coupled to a massless monopole,  
while at $u=-\Lambda^2$ it is coupled to a dyon.  The effective theories
at each singular point are related by an existing $\ZZ_2$ symmetry in the
$u$-plane. This symmetry relates the behavior  of the  theory around one
singularity to its behavior around the other.

At this point it is convenient to recall some facts about the exact
solution of
$N=2$ supersymmetric Yang-Mills  found by Seiberg and Witten \cite{sw} (see
 \cite{luis} for some reviews on this topic and Alvarez-Gaume's lectures \cite{zamora} in this volume).  One of the most important
features of $N=2$ supersymmetric Yang-Mills theory is that its lagrangian
can be  written in terms of a single holomorphic function, the
prepotential 
${\cal F}$.  This prepotential is holomorphic in the sense that it depends
only on the  $N=2$ chiral superfield\footnote{This $N=2$ chiral superfield
should not be  confused with the $N=1$ chiral superfield used in
subsection 4.3.1.}
$\,\,$ $\Psi$ which defines the theory, and not on its complex 
conjugate.  The microscopic theory is defined by a classical quadratic
prepotential:
\be {\cal F}_{{{\rm cl}}}({\Psi})=\half \tau_{{{\rm cl}}}  {\Psi}^2,\qquad 
\tau_{\rm cl}={\theta_{\rm bare}\over2\pi}+{4\pi  i\over g^2_{\rm bare}}.
\label{ccincuentatres}
\ee In terms of this prepotential the lagrangian is given by the following
expression in $N=1$ superspace: 
\be {\cal L}={1\over{4\pi}}{\hbox{\rm Im}}\tr\left[\int d^4\theta  
{\partial{\cal F}(A)\over\partial A}\bar A+\int d^2\theta\half
{\partial^2{\cal F}(A)\over\partial A^2}W^\alpha W_\alpha\right],
\label{ccincuentados}
\ee where $A$ is a chiral $N=1$ superfield containing the fields  
$(\phi,\psi)$, and $W$ is a constrained chiral spinor superfield
containing the  non-abelian gauge field and its $N=1$ superpartner
$(A_\mu,\lambda)$. The lagrangian (\ref{ccincuentados}) is equivalent to
the one entering (\ref{ccuarentacuatro}) after replacing the $N=1$
superfield $\Psi$ there by the
$N=1$ superfield $A$. All the  fields take values in the adjoint
representation of the gauge group, which we take to be $SU(2)$. The
potential term for the complex scalar $\phi$ is: 
\be V(\phi)=\tr\left([\phi,\phi^{\dag}]^2\right).
\label{ccincuentacuatro}
\ee The minimum of this potential is attained at field configurations of
the form $\phi=\half a\sigma^3$, which define the classical moduli space
of  vacua. A convenient gauge invariant parametrization of the vacua is
given by  
$u=\tr\phi^2$, which equals $\half a^2$ semiclassically.  For $u\not =0$,
$SU(2)$ is spontaneously broken to $U(1)$. The spectrum of the theory
splits up  into two massive $N=2$ vector multiplets, which accommodate the
massive
$W^{\pm}$ bosons together with their superpartners, and an $N=2$ abelian
multiplet  which accommodates the $N=2$ photon together with its
superpartners.  For
$u=0$, the full $SU(2)$ symmetry is (classically) restored. 

\begin{figure}                
\centerline{\hskip.4in \epsffile{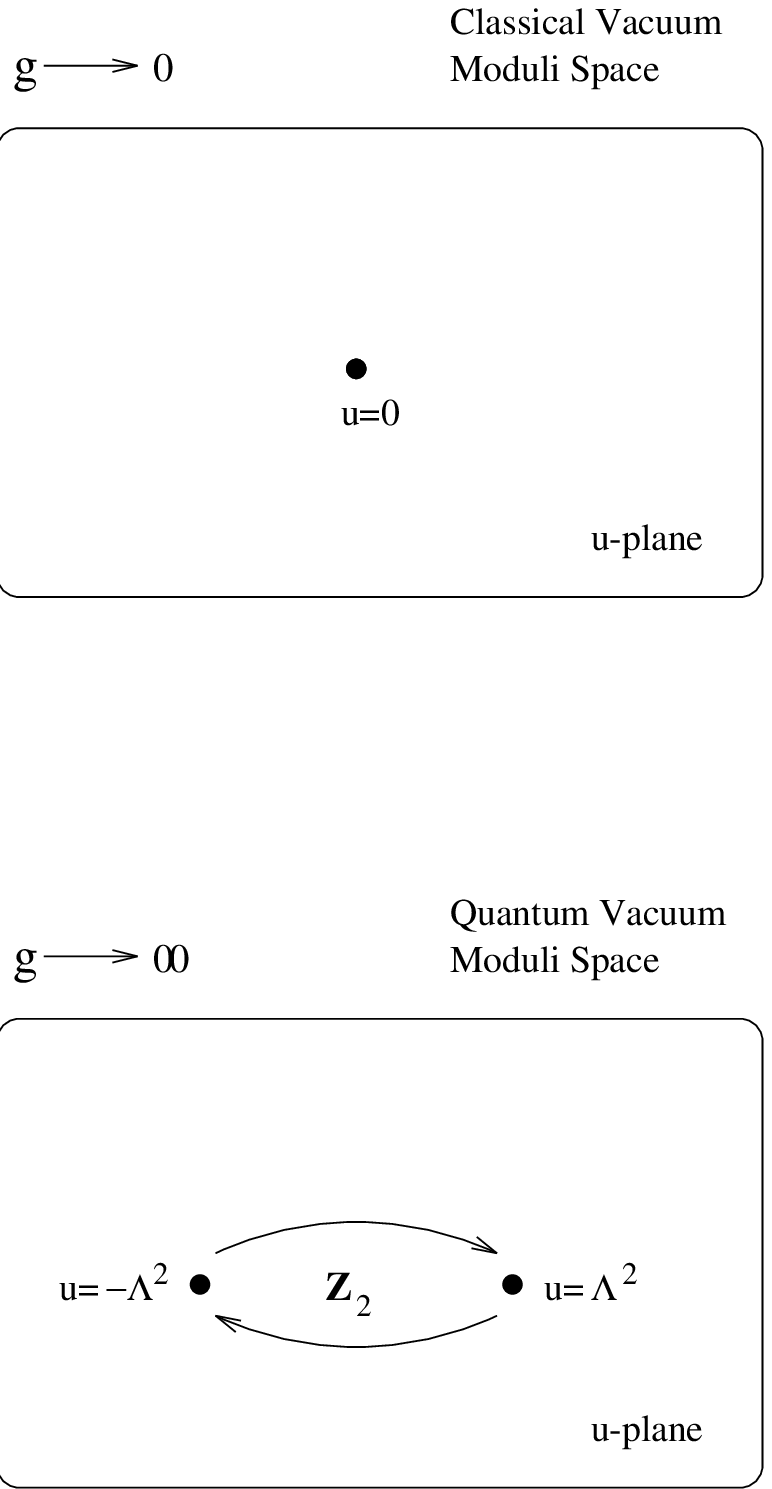}}       
\caption{Classical and quantum moduli spaces.}            
\label{moduli.eps}              
\end{figure}

To study the quantum vacua Seiberg and Witten analyzed the structure of
the low energy  theory, whose effective lagrangian up to two derivatives
is given, after integrating  out the massive modes, by an expression like
(\ref{ccincuentados}) but with a  new effective prepotential depending
only on an abelian multiplet. The  result of their analysis can be
summarized as follows:
\begin{itemize}
\item At the quantum level the $SU(2)$ symmetry is never restored. The 
theory stays in the Coulomb phase throughout the $u$-plane.
\item The moduli space of vacua ($u$-plane) is a complex  one-dimensional
K\"ahler manifold.
\item At the points $u=\pm\Lambda^2$, the prepotential ${\cal F}$ has
singularities.
\item The singularities correspond to the presence of a massless monopole
(at 
$u=\Lambda^2$) and a massless dyon (at $u=-\Lambda^2$).
\item Near each of the singularities the effective action should  include
together with the $N=2$ abelian vector multiplet, a massless monopole or
a dyon hypermultiplet.
\end{itemize} A summary of the main features of both, the classical and the
quantum  moduli spaces, is depicted in Fig.~\ref{moduli.eps}.

\begin{figure}               

 \centerline{\hskip.4in \epsffile{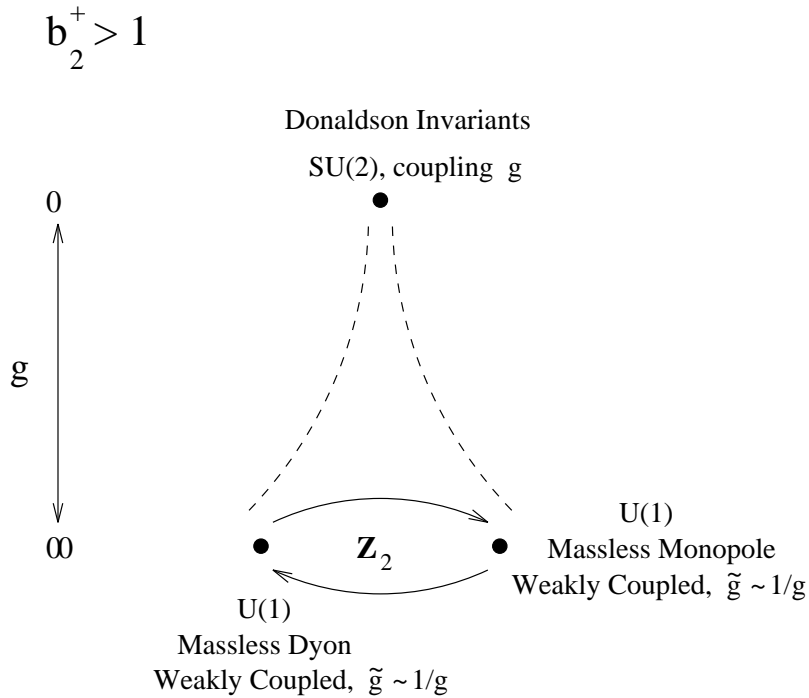}}       

 \caption{Contributions in the strong and weak coupling limits for  
$b_2^+>1$.}            
 \label{u_plane.eps}

\end{figure}

We will describe now the computation of observables. As we did in the
perturbative approach, we will consider the theory  on  manifolds $X$ with
$b^+_2>1$. In the limit $g\to 0$  one has to take into account the
classical moduli space. Since for $b^+_2>1$  there are not abelian
instantons the only contribution come from $u=0$ and one  has to go
through the analysis carried out in our discussion of the perturbative
approach. As described there, one is led to the standard approach to 
Donaldson invariants via integration over the moduli space of non-abelian 
instantons. In the limit  $g\to\infty$, since the supersymmetric theory
is  asymptotically free, we are in the infrared regime, and the
contributions come from  the quantum moduli space.  In the case under
consideration ($b^+_2>1$)  there are no abelian instantons. Since the
abelian gauge field is the only massless field away from the
singularities, the only contributions come from the singular points,
$u=\pm\Lambda^2$, where there are additional massless fields. Near each of
these points, $N=2$ supersymmetry dictates the form of the  weakly coupled
effective theory. Since the observables of the twisted  theory are
independent of the coupling constant, one expects that Donaldson 
invariants can be expressed in terms of vevs of some operators in the
twisted  effective theories around each singular point. This analysis has
been summarized  in Fig.~\ref{u_plane.eps}.

The theory around the monopole singularity is an $N=2$ supersymmetric
abelian gauge theory coupled to a massless hypermultiplet. This theory has
a twisted version which has been constructed in  \cite{rocek,top} from the
point of view of twisting
$N=2$ supersymmetry, and in  \cite{abmono} using the Mathai-Quillen 
formalism. It has been addressed in other works \cite{ans,gian}. The
structure of this theory is similar to the one of Donaldson-Witten 
theory. The resulting action is $\delta$-exact and therefore one can study
the  theory in the weak coupling limit, which, being the theory abelian, corresponds
to the low  energy limit.

Let us describe the structure of the twisted $N=2$ supersymmetric  abelian
gauge theory coupled to a twisted hypermultiplet. We will assume that the
four-dimensional manifold $X$ is a spin manifold. The analysis naturally
extends to the case of manifolds which are not spin as shown in
 \cite{abm}. A hypermultiplet is built out of two chiral
$N=1$ superfields, $Q$ and  $\tilde Q$,
\bea && Q(q^1,\psi_{q\alpha}),\qquad
Q^{\dag}(q^{\dag}_1,\bar\psi_{q\dalpha}),\nonumber\\ && \tilde
Q(q^{\dag}_2,\psi_{\tilde q\alpha}),\qquad \tilde Q^{\dag}  
(q^2,\bar\psi_{\tilde q\dalpha}).
\label{ccincuentacinco}
\eea After the twisting these fields become:
\bea q^i\,\,(0,0,\half)^0 &\too& M^{\alpha}\,\,(\half,0)^0,\nonumber\\
\psi_{q\alpha }\,\,(\half,0,0)^{1} &\too&
\mu_{\alpha}\,\,(\half,0)^{1},\nonumber\\
\bar\psi_{\tilde q\dalpha}\,\,(0,\half,0)^{-1} &\too&
\nu_{\dalpha}\,\,(0,\half)^{-1},\nonumber\\ q^{\dag}_i\,\,(0,0,\half)^{0}
&\too&
\overline M_\alpha\,\,(\half,0)^{0},\\
\bar\psi_{q\dalpha}\,\,(0,\half,0)^{-1} &\too&
\bar\nu_\dalpha\,\,(0,\half)^{-1},\nonumber\\
\psi_{\tilde q\alpha}\,\,(\half,0,0)^{1} &\too&
\bar\mu_{\alpha}\,\,(\half,0)^1.\nonumber
\label{ccincuentaseis}
\eea

The twisted fields  $M_\alpha,
\mu_\alpha$, and $\nu_\dalpha$ belong, respectively, to 
$\Gamma(S^{+}\otimes L)$ and  $\Gamma(S^{-}\otimes L)$, where
$S^{\pm}$ are the positive/negative chirality spin bundles and $L$ is a  
complex line bundle. The action of the twisted abelian effective theory
around the monopole singularity is given by \cite{abmono}:  
\bea S_{\rm AM}&=&\int _{X}\sqrt{g}\big[ g^{ij}D_i\overline M^\alpha D_j
M_\alpha + {1\over 4} R
\overline M^\alpha M_\alpha +{1\over 2} F^{+\alpha\beta}  F_{\alpha\beta}^+
-{1\over 8} \overline M^{(\alpha} M^{\beta)} \overline M_{(\alpha}
M_{\beta)}
\big] \nonumber\\  &&+i\int_{X} \big( \lambda  \wedge *d^{*}d \phi - {1
\over {\sqrt2}}\chi \wedge * p^{+}d \psi +\eta\wedge *d^* \psi \big)  
\nonumber\\ &&+\int_{X}e\Big(i\phi \lambda{\overline M}^{\alpha}M_{\alpha}
+{1  
\over 2{\sqrt2}} {\chi}^{\alpha \beta}({\overline M}_{(\alpha} \mu _{\beta
)}+{\bar
\mu}_{(\alpha} M_{\beta )}) -{i\over 2} (v^{\dot \alpha} D_{\alpha
{\dot\alpha}} {\mu}^{\alpha}-{\bar \mu}^{\alpha} D_{\alpha
{\dot\alpha}}v^{\dot
\alpha}) \nonumber\\ &&\,\,\,\,\,\,\,\,\,\,\,\,\,  -{1 \over 2}({\overline
M}^{\alpha}{\psi}_{\alpha {\dot \alpha}}v^{\dot  
\alpha}-{\bar v}^{\dot \alpha}{\psi}_{\alpha {\dot \alpha}}M^{\alpha})
+{1\over2}  
\eta ({\bar \mu} ^{\alpha} M_{\alpha}-{\overline M}^{\alpha}
\mu_{\alpha})  +{i
\over 4}\phi {\bar v}^{\dot\alpha} v_{\dot\alpha} -\lambda {\bar
\mu}^{\alpha} \mu _{\alpha}\Big).\nonumber\\  
\label{ccincuentasiete} 
\eea This action is invariant under the following scalar symmetry:
\bea  &&[Q,M_{\alpha}]=\mu _{\alpha},\qquad\qquad  \{Q, \mu _{\alpha}  
\}=-i\phi M_{\alpha},\nonumber\\  && \{Q, v_{\dot \alpha} \}=h_{\dot
\alpha},
\qquad\qquad [Q,h_{\dot
\alpha}]=-i \phi v_{\dot \alpha}.
\label{ccincuentaocho}
\eea We only list the transformations for the matter fields, the  
transformations for the rest of the fields are the abelianized version of the ones in
(\ref{cveintenueve}).  The action $S_{\rm AM}$ is $Q$-exact and therefore
the semiclassical approximation is exact.  The main contribution to the
functional integral coming from the bosonic part of  the action is given
by the solutions to the equations:
\begin{equation} F_{\alpha\beta}^+ + \frac{i}{2} \overline M_{(\alpha}
M_{\beta)} =0,
\;\;\;\;\; D_{\alpha\dalpha} M^\alpha = 0.
\label{francesca}
\end{equation}
 These equations  are known as monopole equations \cite{abm}. The tangent
space to the moduli space, ${\cal M}_{\rm AM}$, defined by  these
equations is given by the linearization of (\ref{francesca}), which 
happen to be the field equations:
\bea &&(d\psi)^+_{\alpha\beta}+{i\over2}(\overline  
M_{(\alpha}\mu_{\beta)}+\bar\mu _{(\alpha}M_{\beta)})=0,\nonumber\\
&&D_{\alpha\dalpha}\mu^\alpha+i\psi_{\alpha\dalpha}M^\alpha=0.
\label{ccincuentanueve}
\eea The dimension of the moduli space can be calculated from  
(\ref{ccincuentanueve}) by means of an index theorem, and turns out to
be \cite{abm}, 
\be  d_{{\cal M}_{\rm AM}}=\bigl(c_1(L)\bigr)^2 -\frac{2\chi+3\sigma}{4}.
\label{csesenta}
\ee The only contributions to the partition function come from $d_{{\cal  
M}_{\rm AM}}=0$ (isolated monopoles). Introducing  the shorthand notation,
$x=-2c_1(L)$,  we have: 
\be  d_{{\cal M}_{\rm AM}}=0 \Leftrightarrow x^2=2\chi+3\sigma.
\label{csesentauno}
\ee As in the case of Donaldson-Witten theory, the integration over the  
quantum fluctuations around the background  (\ref{francesca}) gives an  
alternating sum over the different monopole solutions for a given  class
$x$:
\be  n_x=\sum_i \epsilon_{i,x},\qquad \epsilon_{i,x}=\pm 1.
\label{csesentados}
\ee The $n_x$ are the partition functions of the twisted abelian theory
for a fixed class $x$ (compare to (\ref{vanessa})). Those classes such that
(\ref{csesentauno}) holds and
$n_x\ne 0$ are  called {\it basic classes}. The quantities $n_x$ turn out
to constitute a new set of topological invariants for four-manifolds known
as Seiberg-Witten invariants. 

To fix ideas let us analyze in certain detail the outline of the 
calculation of the partition function of Donaldson-Witten theory on a
manifold $X$  with
$b_2^+>1$ and for gauge group $SU(2)$. Recall that we are dealing with  a
TQFT which corresponds to a twisted version of $N=2$ supersymmetric 
Yang-Mills theory. In the weak coupling limit, $g\to 0$, the partition
function is dominated by
$SU(2)$ instanton configurations as in (\ref{vanessa}):
\be
Z=\sum_{k=0}^{\infty}\delta(\underbrace{8k-{3\over2}(\chi+\sigma)}_{d_{{ 
\cal M}_{\rm ASD}}})Z_k,
\label{csesentacuatro}
\ee where,
\be Z_k=\sum_{ {\scriptstyle {\rm solutions}}\atop
\scriptstyle i}(-1)^{\nu_i},\qquad \nu_i=0,1.
\label{csesentacinco}
\ee

In the strong-coupling limit,
$g\to\infty$, or,
$\tilde g={1/ g}\to 0$, by analogy with the physical theory,  we expect 
that the correct description is given by a sum over the partition
functions of effective TQFTs which are twisted versions of the
corresponding effective description of the physical theory at the points
$u=\pm\Lambda^2$:
\be Z=c(Z_{u=\Lambda^2}+Z_{u=-\Lambda^2}),
\label{csesentasiete}
\ee where $c$ is a factor to be fixed. The partition functions of the
twisted theories are dominated by configurations satisfying the monopole
equations  (\ref{francesca}) for classes $x$ satisfying
(\ref{csesentauno}). For
$Z_{u=\Lambda^2}$ we have:
\be  Z_{u=\Lambda^2}=\sum_x n_x\delta(x^2-2\chi-3\sigma),\qquad x=-2c_1(L),
\label{csesentaocho}
\ee  with the Seiberg-Witten invariants given by 
\be  n_x=\sum_{{\scriptstyle {\rm solutions}}\atop \scriptstyle i}
(-1)^{\mu_i},\qquad \mu_i=0,1.
\label{csesentanueve}
\ee

The partition function at the dyon singularity $Z_{u=-\Lambda^2}$  is
related to the previous one by a $\ZZ_2$ transformation. This  transformation  is
anomalous on a gravitational background, and therefore there is  a
contribution from the measure when  comparing
$Z_{u=\Lambda^2}$ and $Z_{u=-\Lambda^2}$. We will discuss the details  of
this issue below. Now we take for granted that the relation between both 
partition functions is given by:
\be Z_{u=-\Lambda^2}=i^{{\chi+\sigma\over4}}Z_{u=\Lambda^2}.
\label{csetenta}
\ee 

Being the theory topological, the result obtained in both limits should 
be the same. We then obtain the following relation:
\be Z=\sum_{k=0}^{\infty}\delta({8k-{3\over2}(\chi+\sigma)})
Z_k=c\sum_{\scriptstyle  x\atop{{\scriptstyle {\rm basic}}\atop
{\scriptstyle {\rm classes}}}}\delta(x^2-2\chi-3\sigma)
\left[n_x+i^{{\chi+\sigma\over4}}n_x\right].
\label{csetentauno}
\ee The quantity $c$ is fixed comparing both sides of this equation  for
different four-manifolds $X$ with $b_2^+>1$. It turns out that,
\begin{equation} c= 2^{1+\frac{1}{4}(7\chi+11\sigma)}.
\label{lace}
\end{equation} This quantity should be computable from field-theoretical
arguments  but, to our knowledge, it is not known at the moment how to do
it. Some steps to determine it have been given in reference  \cite{sdual}.

Equation (\ref{csetentauno}) constitutes a field-theory prediction  which
has been verified on all manifolds in which it has been tested. Its
content  is very important because it relates topological information
coming from  two apparently unrelated moduli spaces. On the left hand side
the  contributions  are given by non-abelian instanton configurations.
However, on the right hand side the contributions come from abelian
monopole configurations. It is also important to remark that  the fact
that eq. (\ref{csetentauno}) holds constitutes a very important test for 
Seiberg-Witten theory. Indeed, for example, if the quantum moduli space for
$SU(2)$  had had a number of singularities different from two, the right hand
side of  (\ref{csetentauno}) would have been different, spoiling the
agreement.

The partition function is not the only observable that can be computed  by
making use of Seiberg-Witten theory. One can indeed compute the full  
generating function (\ref{ccuarentaocho}). The steps needed to carry out 
this computation are the following:
\begin{enumerate}
\item Work out the form of the observables in the variables of the 
effective theory around the monopole singularity, $u=\Lambda^2$.
\item Work out the contribution from the dyon singularity,  
$u=-\Lambda^2$, using the $\ZZ_2$ symmetry present in the $u$-plane.
\item Sum over all basic classes $x$. 
\end{enumerate} We will go now through these steps in turn.

\vspace{1pc}
\noindent
$1$. As in the case of the abstract approach, let us assume that $X$  is
simply connected. Recall that in this case the relevant observables are 
(\ref{ccuarentacinco}). We reproduce them here:
\bea  {\cal O}&=&\frac{1}{8\pi^2}\tr(\phi^2),\nonumber\\ 
I(\Sigma_a)&=&\frac{1}{4\pi^2}\intl_{\Sigma_a} \tr(\phi  
F+\half\psi\wedge\psi),
\label{otravez}
\eea being $\{\Sigma_a\}_{a=1,\ldots,b_2(X)}$ a basis of $H_2(X)$. These
observables are the ingredients of the generating function
(\ref{ccuarentaocho}),
\be
\left\langle\exp\left({\sum_a\alpha_a I(\Sigma_a)+\lambda{\cal  O}}\right)
\right\rangle,
\label{csetentatres}
\ee which is the goal of our computation. Recall that $\lambda$ and  
$\alpha_a$ are arbitrary parameters.  

In computing (\ref{csetentatres}) we must address the question of what  is
the form of the observables of Donaldson-Witten theory in terms of 
operators of the effective abelian theory. To answer this question we will
use the expansion of the observables in the untwisted, physical theory,
together  with the descent equations in the topological abelian theory. We
follow the argument presented in  \cite{puri} which is originally due to
Witten. The descent equations for the abelian monopole theory can be found
in  
 \cite{abmono}. Near the monopole singularity, the $u$ variable has  the  
expansion \cite{sw}:
\be u(a_D)= \Lambda^2 + \Big( {du \over da_D} \Big)_{0}a_D + {\rm  higher
\,\  order}, 
\label{expan}
\ee where $(du / da_D)_{0}= -2i \Lambda$, while ``higher order" stands for
operators of  higher dimensions in the expansion. The field $a_D$
corresponds to the  field 
$\phi_D$ of the topological abelian theory \cite{abmono}, while the 
gauge-invariant parameter $u$ corresponds to the observable 
(\ref{raquela}). In  terms of observables of the corresponding twisted
theories, the  expansion  (\ref{expan}) reads,
\be {\cal O}= \langle {\cal O} \rangle -{1 \over \pi} \langle V \rangle  
\phi_D  + {\rm higher \,\ order}, 
\label{descef}
\ee where $\langle {\cal O} \rangle$, $\langle V \rangle$ are real  
$c$-numbers  which should be related to the values of $u(0)$, $(du /
da_D)_{0}$ in  the  untwisted theory. From the observable ${\cal O}$ one
can obtain the observable 
${\cal O}^{(2)}$ by the descent procedure. By applying it in the  abelian
TQFT to the right hand side of (\ref{descef}), we obtain:
\be {\cal O}^{(2)}=-{1\over\pi}\langle V\rangle F_D + {\rm higher \,\ 
order},
\label{descedos}
\ee where $F_D$ is the dual electromagnetic field strength  (associated to
the  magnetic  monopole). In particular, taking into account that  
$x=-c_1(L^2)=-[F_D]/\pi$,  where $[F_D]$ denotes the cohomology class of
the two-form $F_D$, we  finally obtain:
\be I(\Sigma)=\langle V\rangle(\Sigma \cdot x) + {\rm higher \,\ order},
\label{final}
\ee where the dot denotes the pairing between $2$-cohomology and  
$2$-homology. From the point of view of TQFT, higher  dimensional terms
should not contribute to the expansion  because of  the invariance of the
theory under rescalings of the metric.  Notice that  the lower order terms
in (\ref{descef}) and (\ref{final}) are $c$-numbers, \ie,  they have
zero-ghost numbers. This means that the only contributions to 
(\ref{csetentatres}) will come from configurations such that the 
dimension of the moduli space of abelian monopoles,
$d_{{\cal M}_{\rm AM}}$, vanishes.  This dimension is given in
(\ref{csesenta}). Thus the only classes  contributing to
(\ref{csetentatres}) are again the basic classes, \ie,  classes, $x$,
satisfying $x^2=2\chi + 3\sigma$ and $n_x\ne 0$. Manifolds for which  this
holds are called of {\it simple type}. 

We are now in the position to evaluate the correlation function
(\ref{csetentatres}) at the monopole singularity. This is rather  simple
since  (\ref{descef}) and (\ref{final}) are $c$-numbers. Only additional  
contact terms for the operators $I(\Sigma)$ appear (see
reference \cite{wijmp} for a discussion  on this point). These terms have
the following form. Let  us introduce $v=\sum_a \alpha_a I(\Sigma_a)$,
then the contribution turns out to be \cite{wijmp}:
$\gamma v^2 = \gamma \sum_{a,b}\alpha_a \alpha_b (\Sigma_a,
\Sigma_b)$, where $\gamma$  is a real number.  Taking into account this
term and eq. (\ref{descef}) and (\ref{final})  one finds that the
contribution at the monopole singularity is:
\be C_1 \hbox{\rm exp}(\gamma v^2 +\lambda \langle {\cal O} \rangle)  
\sum_{x} n_{x} {\rm e}^  {\langle V \rangle v \cdot x},
\label{singu}
\ee where  $C_1$ is the factor $c$ which appeared in (\ref{csesentasiete})
and turned out to be (\ref{lace}).



%
\vspace{1pc}

\noindent 2. Next, we work out the contribution from the dyon
singularity,  
$u=-\Lambda^2$.  This contribution is related to the one from
$u=\Lambda^2$ by a  
$\ZZ_2$ transformation, which is the anomaly-free symmetry on the
$u$-plane which remains after the breaking of the  chiral symmetry
$U(1)_{\cal R}$. Let us  begin by  recalling the transformations of the
fields entering the observables  under the $U(1)_{\cal R}$-transformations:
\bea  &&\psi^1_\alpha\too \ex^{-i\varphi}\psi^1_\alpha,\nonumber\\
&&\psi^2_\alpha\too \ex^{-i\varphi}\psi^2_\alpha,\nonumber\\ &&B\,\too
\ex^{-2i\varphi}B.\nonumber
\label{csetentasiete}
\eea Instanton effects break this symmetry down to ${\ZZ}_8$ ($4N_c-2N_f$
in the general case of $SU(N_c)$ gauge group with $N_f$ hypermultiplets
in  the fundamental representation). Under this anomaly-free ${\ZZ}_8$, 
\be B\too \ex^{-2i({2\pi\over8})}B= \ex^{-{i\pi\over2}}B,
\label{csetentaocho}
\ee and therefore,
\be  u=\tr(B^2)\too\ex^{-i\pi}u=-u,
\label{csetentanueve}
\ee which gives a ${\ZZ}_2$ symmetry on the $u$-plane. This ${\ZZ}_2$ 
symmetry relates the contributions to the vevs from $u=\Lambda^2$ to the
ones  from
$u=-\Lambda^2$. Under the $\ZZ_8$ symmetry, the observables 
(\ref{ccuarentacinco}) transform as follows:
\bea I(\Sigma_a)=\frac{1}{4\pi^2}\intl_{\Sigma_a}\tr\left(\phi
F+\half\psi\wedge\psi\right)&\too&
\ex^{-{i\pi\over2}}I(\Sigma_a)=-iI(\Sigma_a),\nonumber\\  {\cal
O}=\frac{1}{8\pi^2}\tr(\phi)^2&\too& \ex^{-i\pi}{\cal O}=-{\cal
O},\nonumber\\
\label{cochenta}
\eea hence, using (\ref{singu}) one finds:
\bea u=\Lambda^2,\qquad&& C_1\exp{\displaystyle{\left(\gamma
v^2+\lambda\langle{\cal O}\rangle +
\langle V\rangle v\cdot x\right)}},\nonumber\\ u=-\Lambda^2,\qquad&&
C_2\exp{\displaystyle{\left(-\gamma v^2-\lambda\langle{\cal O}\rangle 
-i\langle V\rangle v\cdot x\right)}}.
\label{cochentauno}
\eea

The quantities $C_2$ and $C_1$ are  related because on a curved 
background the ${\ZZ}_8$ transformation, while being preserved by gauge 
instantons, picks anomalous contributions from the measure due to
gravitational  anomalies. The contribution is of the form
$\exp{{i\pi\over2}\Delta}$, where
$\Delta$  is the index of the Dirac operator. For a basic class, ${\rm
dim}\,{\cal M}_{\rm AM}=0$, and therefore, from (\ref{csesenta}), 
$(c_1(L))^2={2\chi+3\sigma\over4}$. The index of the Dirac operator is,  
\be
\Delta=-{\sigma\over8}+\half (c_1(L))^2={\chi+\sigma\over4}\in {\ZZ},
\label{cochentatres}
\ee and therefore the anomaly can be written as $i^{\Delta}$, with  
$\Delta= {\chi+\sigma\over4}$. Then,
\be C_2=i^\Delta C_1.
\label{cochentacuatro}
\ee
\vspace{1pc}

\noindent 3. Finally, we take both contributions and sum over basic
classes. The final form of the generating function of vevs of observables
turns out to be:
\bea
\langle \ex^{\scriptstyle{\left(\sum_a \alpha_a  I(\Sigma_a)+\lambda{\cal
O}\right)}}\rangle =C_1\left[\ex^{\scriptstyle{\left(\gamma 
v^2+\langle{\cal O}\rangle \lambda\right)}}
\sum_x n_x\ex^{\scriptstyle{\langle V\rangle v\cdot x}}+i^\Delta  
\ex^{\scriptstyle{\left(-\gamma v^2-\langle{\cal O}\rangle  
\lambda\right)}}
\sum_x n_x\ex^{\scriptstyle{-i\langle V\rangle v\cdot 
x}}\right].\nonumber\\
\label{cochentacinco}
\eea By comparing to known results by Kronheimer and Mrowka \cite{km}  the  
unknown constants in (\ref{cochentacinco}) are fixed to be:
\be
\gamma =\half,\qquad \langle{\cal O}\rangle=2,\qquad
\langle{V}\rangle=1,\qquad
C_1=2^{\scriptstyle{1+{1\over4}(7\chi+11\sigma)}}.
\label{cochentaseis}
\ee The ratio between $\langle{V}\rangle$ and $\langle{\cal O}\rangle$ is
predicted by Seiberg-Witten theory since both originated from eq.
(\ref{expan}). It agrees with (\ref{cochentaseis}). Notice  that  these
constants, $\gamma$,
$\langle{V}\rangle$, and $\langle{\cal O}\rangle$, coming from  the
structure of the physical theory in $\RR^4$, should be universal, \ie,
entirely independent of the manifold $X$. This turns out to be the  case
according to the values (\ref{cochentaseis}), a very important test of 
our arguments. Different aspects of Seiberg-Witten solution are reflected 
in  (\ref{cochentacinco}). The fact that this formula fits all known 
mathematical results for simply-connected manifolds with $b_2^+>1$ is
rather  satisfactory from the physical point of view.

Gathering all the preceding results we obtain the final expression for  the
generating function of Donaldson invariants:
\bea
\langle \ex^{\scriptstyle{\left(\sum_a \alpha_a  I(\Sigma_a)+\lambda{\cal
O}\right)}}\rangle=2^{\scriptstyle{1+{1\over4}(7\chi+11\sigma)}}
\left[\ex^{\left({{v^2}\over2}+2\lambda\right)}
\sum_x n_x\ex^{\scriptstyle{v\cdot x}}+i^\Delta  
\ex^{\left(-{v^2\over2}-2\lambda\right)}
\sum_x n_x\ex^{\scriptstyle{-iv\cdot x}}\right].\nonumber\\
\label{cochentasiete}
\eea The expression above verifies the so-called {\it simple type}
condition:
\be
\left({\partial^2\over{\partial\lambda^2}}-4\right)
\langle \ex^{\scriptstyle{\left(\sum_a \alpha_a  I(\Sigma_a)+\lambda{\cal
O}\right)}}\rangle=0.
\label{cochentaocho}
\ee All simply-connected four-manifolds with  
$b^{+}_2>1$ for which (\ref{cochentasiete}) is known verify this property. 

\vfill
\newpage

\section{Generalized Donaldson-Witten Theory}
\setcounter{equation}{0}

So far we have discussed two different moduli problems in  four-dimensional
topology, one defined by the ASD instanton  equations and another one
defined by the Seiberg-Witten monopole  equations. There is a natural
generalization of these moduli  problems which involves a non-abelian
gauge group and also  includes spinor fields. It is the moduli problem
defined by  the {\it  non-abelian monopole equations}, introduced in
reference \cite{nabm}  in the context of the  Mathai-Quillen formalism and
as a generalization of Donaldson theory. It has been also considered in
reference \cite{park,ans}, as well as in the  mathematical
literature \cite{oko,tele,pt,oscar}. 

In order to introduce these equations in the case of $G=SU(N)$ and the 
monopole fields in the fundamental representation ${\bf N}$ of $G$,  let
us consider a  Riemannian four-manifold $X$ together  with a principal
$SU(N)$-bundle $P$ and a vector bundle $E$ associated to $P$  through the
fundamental representation.  Suppose for simplicity that the manifold is  
spin, and consider a section $M_{\alpha}^i$ of $S^{+}\otimes  E$. The
non-abelian monopole equations read in this case: 
\bea && F^{+ij}_{\alpha \beta} + i ( {\overline M}^j_{(\alpha}
M^i_{\beta)}- {\delta^{ij} \over N}{\overline M}^k_{(\alpha} M^k_{\beta)})
 =0, 
\nonumber\\  && (D^{\dot \alpha \alpha}_{E}M_{\alpha})^i = 0.
\label{namon}
\eea Starting from these equations it is possible to build the associated  
topological field theory within the Mathai-Quillen formalism. Not
surprisingly, the  resulting theory is the non-abelian version of the
topological theory of abelian  monopoles, that is, a twisted version of
$N=2$ super Yang-Mills coupled to one  massless hypermultiplet. The field
content is just the  non-abelian  version of that of the abelian monopole
theory. In addition to the fields in Donaldson-Witten theory, we have the
following matter fields:
\be M_\alpha,\mu_\alpha\in\Gamma(S^+\otimes E),\qquad \nu_\dalpha\in
\Gamma (S^-\otimes E),
\label{cochentanueve}
\ee together with their corresponding complex conjugates. The action for
the model takes the form:
\bea S_{\rm NAM}&=&
\int_{X} e \big[ g^{\mu\nu}D_\mu\overline M^\alpha D_\nu M_\alpha + 
{1\over 4} R
\overline M^\alpha M_\alpha \nonumber\\ & & \,\,\,\,\,\,\,\,\ -{1\over  4}
\tr (F^{+\alpha\beta} F_{\alpha\beta}^+)  +{1\over 4}(\overline 
M^{(\alpha} T^a M^{\beta)}) (\overline M_{(\alpha} T^a
M_{\beta)})]\nonumber\\
 &+& \int_X \tr\big({ i\over 2} \eta \wedge *d_A^{*} \psi +{i \over 
2{\sqrt 2}}\chi ^{\alpha
\beta}(p^+(d_A\psi))_{\alpha\beta} +{i\over
8}\chi^{\alpha\beta}[\chi_{\alpha\beta},\phi]\nonumber\\ & &
\,\,\,\,\,\,\,\,\ + {i
\over 2} \lambda \wedge *d_A^{*}d_A \phi - { 1\over 2}
\lambda\wedge *[*\psi,\psi] \big) 
\nonumber\\  &+&\int_{X}e\Big({i\over 2}{\overline M}^{\alpha}\{\phi,
\lambda\} M_{\alpha}  -{1
\over {\sqrt2}} ({\overline M}_{\alpha} {\chi}^{\alpha \beta} \mu  _{\beta
}-{\bar
\mu}_{\alpha}{\chi}^{\alpha \beta} M_{\beta}) \nonumber\\ & &
\,\,\,\,\,\,\,\,\ -{i\over 2} ({\overline v}_{\dot
\alpha} D^{{\dot\alpha} \alpha } {\mu}^{\alpha}+{\bar \mu}^{\alpha} 
D_{\alpha {\dot\alpha}} v^{\dot \alpha}) +{1 \over 2}({\overline
M}^{\alpha}{\psi}_{\alpha {\dot \alpha}}v^{\dot \alpha}+{\bar v}_{\dot
\alpha}{\psi}^{{\dot \alpha}\alpha }M_{\alpha}) \nonumber\\ & &
\,\,\,\,\,\,\,\,\ -{1\over2}  ({\bar
\mu}^{\alpha}  \eta M_{\alpha}+{\overline M}^{\alpha}
 \eta\mu_{\alpha}) +{i \over 4}{\bar v}^{\dot\alpha} \phi  v_{\dot\alpha}
-{\bar
\mu}^{\alpha} \lambda \mu _{\alpha}\Big).
\label{cnoventa}
\eea This action can be derived either by applying the Mathai-Quillen
formalism,  as we discuss below, or by twisting the corresponding action
for the  physical $N=2$ supersymmetric theory. However, as it happens in
Donaldson-Witten theory,  the action which comes directly from the 
twisting and that in (\ref{cnoventa}) differ by the term (\ref{rourke}),
which, being of the form  $\sim \{Q,\int\eta[\phi,\lambda]\}$,  can be 
safely ignored.  

From the monopole eq. (\ref{namon}) follows that the appropriate 
geometric setting is the following. The field space is ${\cal
A}\times\Gamma(X,S^+\otimes E)$, which is  the space of gauge connections
on
$P$ and positive chirality spinors in the  representation  ${\bf N}$ of
$G$. The vector bundle has as  fibre, ${\cal F}=\Omega^{2,+}(X,\ad
P)\oplus\Gamma(X,S^+\otimes E)$, as dictated by the  quantum numbers of the
monopole equations. These equations  are arranged into a section of the
vector bundle $({\cal A}\times\Gamma(X,S^+\otimes E))\times  {\cal F}$:
\be s(A,M)=\Big({1 \over \sqrt 2} \big(F^{+ ij}_{\alpha \beta}+i 
({\overline M}_{(\alpha}^j  M_{\beta )}^i-{\delta^{ij}\over N} {\overline 
M}_{(\alpha}^k  M_{\beta )}^k)\big), (D^{{\dot \alpha}\alpha
}M_{\alpha})^i\Big),  
\label{cnoventauno}
\ee  in such a way that the zero locus of this section gives precisely the 
desired moduli space. The action (\ref{cnoventa}) is exact with respect 
to the following transformations:
\be  
\begin{array}{cclcccl}  [Q,A] &=& \psi,&
\,\,\,\,\,\,\,\,\,\,\,\,\,\,\,\,\,\,\,\,\,\,\,\,\,\,\,\ &  
[Q,M_{\alpha}^i] &=&\mu_{\alpha}^i,  \nonumber\\
\{ Q,\psi \} &=& d_A \phi,&
\,\,\,\,\,\,\,\,\,\,\,\,\,\,\,\,\,\,\,\,\,\,\,\,\,\,\,\ &
\{Q, \mu_{\alpha}^i \} &=& -i\phi^{ij} M_{\alpha}^j, \nonumber \\ 
{[}{Q},{\phi}{]} &=&0, &
\,\,\,\,\,\,\,\,\,\,\,\,\,\,\,\,\,\,\,\,\,\,\,\,\,\,\,\ &
 \{Q, v_{\dot \alpha}^i \} &=&h_{\dot \alpha}^i, \nonumber\\
\{ Q,\chi_{ \mu\nu} \} &=&H_{\mu\nu},&
\,\,\,\,\,\,\,\,\,\,\,\,\,\,\,\,\,\,\,\,\,\,\,\,\,\,\,\ & {[}  {Q},{h_{\dot
\alpha}^i} {]}&=&-i \phi^{ij} v_{\dot \alpha}^j, \nonumber\\  {[}
{Q},{H_{\mu\nu}}{]} &=&i[\chi_{\mu\nu},\phi ],& 
\,\,\,\,\,\,\,\,\,\,\,\,\,\,\,\,\,\,\,\,\,\,\,\,\,\,\,\ &
 \{Q, \eta \} &=&i[\lambda,\phi], \nonumber\\  {[}Q, \lambda] &=& \eta. &
\,\,\,\,\,\,\,\,\,\,\,\,\,\,\,\,\,\,\,\,\,\,\,\,\,\,\,\ & 
\end{array}
\label{cnoventados}
\ee  The gauge fermions of the theory are:
\bea
\Psi _{\rm loc} &=&
\int _{X} e \Big[{i\over 2} \chi^{\alpha \beta ji}\Big({1  
\over {\sqrt 2}}
\big(F^{+ij}_{\alpha \beta} +i ({\overline M}_{(\alpha}^j M_{\beta)}^i
-{\delta^{ij}\over N}{\overline M}_{(\alpha}^k M_{\beta)}^k)\big) - {i  
\over 4} H_{\alpha \beta}^{ji} \Big)
\nonumber\\  & & \,\,\,\,\,\,\,\,\,\,\,\,\,\,\,\,\,\,\,\,\,\,\,\,\,
 -{i\over 2} ({\bar v}_{\dot \alpha}D^{\dot \alpha \alpha }
M_{\alpha}-{\overline M}^{\alpha}D_{\alpha \dot \alpha}v^{\dot  
\alpha})-{1
\over 8} ({\bar v}_{\dot \alpha} h^{\dot \alpha}+{\bar h}_{\dot
\alpha}v^{\dot \alpha} ) \Big],
\nonumber\\  
\Psi _{\rm proj}&= & -{ 1\over 2}\int _{X} \big[i
 \tr(\lambda \wedge *d_A^{*} \psi) +e  ({\bar
\mu}^{\alpha}\lambda M_{\alpha}- {\overline M}^{\alpha} \lambda
\mu_{\alpha}) \big].
\label{cnoventacuatro} 
\eea

The dimension of the moduli space of non-abelian monopoles is given  by the
dimension of the corresponding tangent space, which in turn  is defined by
the linearized version of the monopole equations:
\bea &&p^+(d_A\psi)_{\alpha
\beta}^{ij}+ {i\over2}\big({\overline M}_{(\alpha}^j \mu_{\beta  )}^i+{\bar
\mu}_{(\alpha}^j M_{\beta )}^i -{\delta^{ij}\over N}({\overline
M}_{(\alpha}^k
\mu_{\beta )}^k+{\bar
\mu}_{(\alpha}^k M_{\beta )}^k)=0,\nonumber\\ & &(D^{{\dot \alpha}\alpha}
\mu_{\alpha})^i+i\psi^{{\dot
\alpha}\alpha}_{ij}M_{\alpha}^j =0. \label{cnoventaseis}
\eea  The dimension of the moduli space counts essentially the number of 
independent solutions to these equations modulo gauge transformations, 
and it is given by a suitable index theorem, with the result:
\bea {\rm dim} \,\ {\cal M}_{\rm NA}&=&{\rm dim} \,\ {\cal M}_{\rm  ASD}+2
\,\ {\rm index}\,\ D_E\nonumber\\ &=&(4N-2)c_2(E)-{N^2-1 \over 2}(\chi
+\sigma)-{N
\over 4}\sigma,
\label{cnoventasiete}
\eea  Notice that ${\cal M}_{\rm ASD}\subset{\cal M}_{\rm NA}$. In
addition to this, the usual conditions to have a well-defined moduli
problem (like the reducibility) are essentially the same  as in Donaldson
theory. 

The observables of the theory are the same as in Donaldson-Witten  theory
since no non-trivial  observables involving matter fields have been found.
The  topological  invariants are then given by correlation functions of
the form  (\ref{clara}):
\begin{equation}
\langle {\cal O}^{(k_1)} {\cal O}^{(k_2)} \cdots {\cal O}^{(k_p)} \rangle =
\int  {\cal O}^{(k_1)} {\cal O}^{(k_2)} \cdots {\cal O}^{(k_p)} \exp
({-{S}/{g^2}}),
\label{aclarados}
\end{equation}

In the perturbative regime, $g\to 0$, one finds the same pattern as in  
ordinary Donaldson-Witten theory. There is a map like in (\ref{prodi}),
$ H_k(X)\too H^k({\cal M}_{\rm NA}) $, which implies that the vevs of the
theory provide a new set of  polynomials in
$H_{k_1}(X)\times H_{k_2}(X)\times\ldots\times H_{k_p}(X)$. As in  the
case of ordinary Donaldson-Witten theory, the perturbative approach does
not  provide any further insight into the precise form of these
topological  invariants. Fortunately, it is again possible to apply the
results of Seiberg and  Witten  on $N=2$ supersymmetric theories to
analyze the model at hand in the non-perturbative regime, in much the same
way as it has been done in the  case of  Donaldson-Witten theory.

\begin{figure}                
 \centerline{\hskip.4in
\epsffile{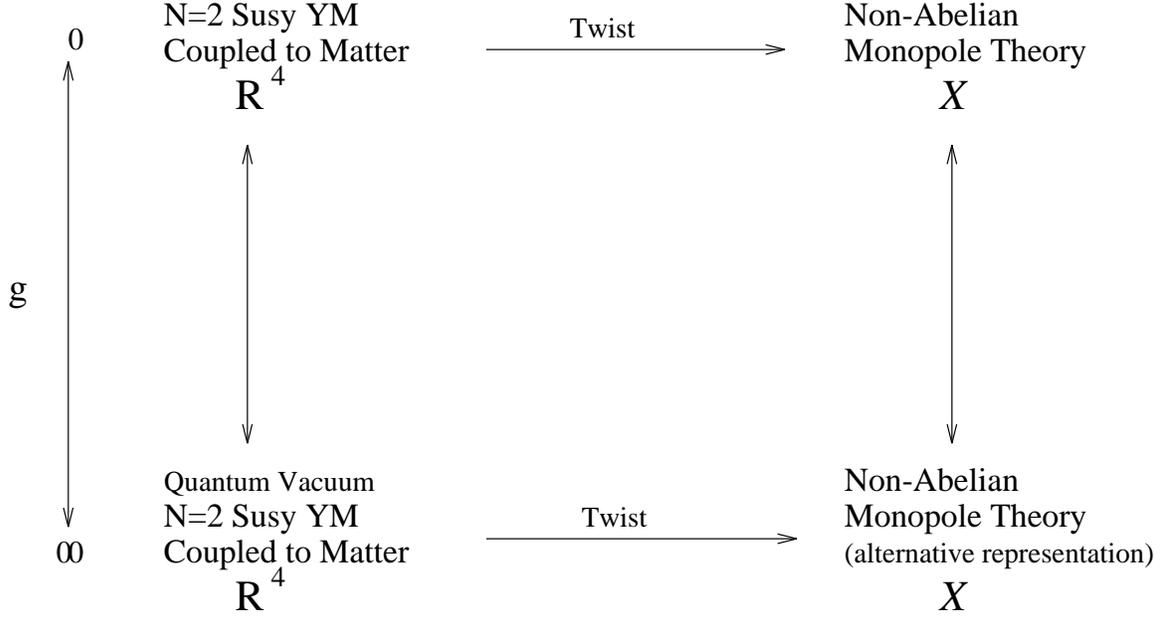}}       
 \caption{Schematic diagram of the concrete approach for non-abelian
monopoles.}            
\label{namonop.eps}             
\end{figure}

We will now discuss the non-perturbative approach. We will follow the same
strategy as in the case of Donaldson-Witten theory. This has been depicted
in  Fig.~\ref{namonop.eps}. The physical theory underlying the theory of
non-abelian monopoles  is an 
$N=2$ supersymmetric Yang-Mills theory coupled to one massless
hypermultiplet in the fundamental representation of the gauge group, which
we take to be  
$SU(2)$.  This theory is asymptotically free. Hence, it is weakly coupled 
($g\to 0$) in the ultraviolet, and strongly coupled ($g\to
\infty$) in  the infrared. The infrared behavior of this theory has been 
also determined by Seiberg and Witten \cite{sw}. Their results can  be
summarized as follows:
\begin{itemize}
\item The quantum moduli space of vacua is a one-dimensional complex
K\"ahler manifold (the $u$-plane). 
\item For any $u$ there is an unbroken $U(1)$ gauge symmetry (Coulomb 
phase).
\item At a generic point on the $u$-plane the only light degree of 
freedom is the $U(1)$ gauge field (together with its $N=2$ superpartners). 
\item There are three singularities at finite values of $u$.
\item Near each of these singularities  a magnetic monopole or dyon 
becomes massless and weakly coupled to a dual $U(1)$ gauge field.
\end{itemize} A scheme of the quantum moduli space is presented in Fig.
$5$.

\begin{figure}                
 \centerline{\hskip.4in
\epsffile{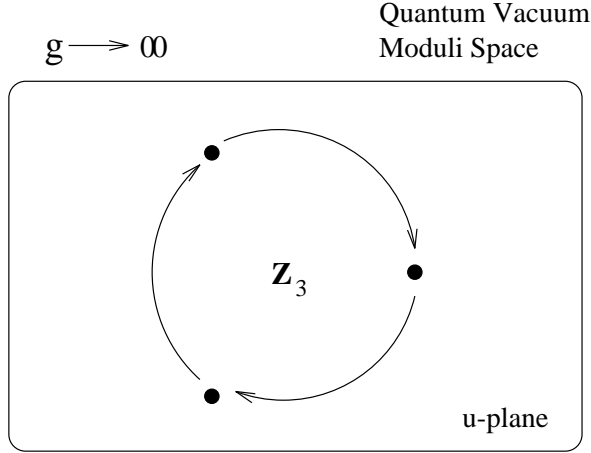}}       
 \caption{The quantum moduli space for
$N=2$ supersymmetric Yang-Mills theory with
$SU(2)$ as gauge group, coupled to a hypermultiplet in the fundamental
representation.}            
\label{triality.eps}              
\end{figure}               

For $X$ such that $b^+_2>1$ (there are no abelian instantons) the only 
contributions come from the three singularities. Following the same
arguments  as in the abelian case one finds the following general result:
\be
\langle \ex^{\scriptstyle{\left(\sum_a \alpha_a  I(\Sigma_a)+\lambda{\cal
O}\right)}}\rangle=\sum_{i=1}^3
C_i\ex^{\scriptstyle{\left({\eta_i\over2}v^2+\xi_i\lambda\right)}}
\sum_x n_x\ex^{\scriptstyle{\zeta _iv\cdot x}}.
\label{cnoventanueve}
\ee Relations among the quantities $C_i$, $\eta_i$, $\zeta_i$ and $\xi_i$ 
are obtained by using the  broken $U(1)_{\cal R}$ symmetry. Recall that:
\bea
\psi^1_\alpha&\too&\ex^{-i\varphi}\psi^1_\alpha,\nonumber\\
\psi^2_\alpha&\too&\ex^{-i\varphi}\psi^2_\alpha,\nonumber\\
B\,&\too&\ex^{-2i\varphi}B\,,\nonumber\\
\psi_{qi}&\too&\ex^{i\varphi}\psi_{qi},\nonumber\\
\psi_{\tilde qi}&\too&\ex^{i\varphi}\psi_{\tilde qi}.
\label{doscientos}
\eea Instanton effects break this symmetry down to ${\ZZ}_6$. Under this  
${\ZZ}_6$,
\be B\too \ex^{-{2i\pi\over3}}B,
\label{duno}
\ee and therefore 
\be  u=\tr(B^2)\too\ex^{-{4i\pi\over3}}u=\ex^{{2i\pi\over3}}u,
\label{ddos}
\ee which generates a ${\ZZ}_3$ symmetry on the $u$-plane which 
interchanges the three singularities. Under this symmetry  the observables
transform as follows:
\bea &&I(\Sigma_a)\too  
\ex^{{-2i\pi\over3}}I(\Sigma_a),\nonumber\\  &&{\cal O}\too
\ex^{{2i\pi\over3}}{\cal O}.
\label{dtres}
\eea This implies for the unknown constants in (\ref{cnoventanueve})  the
following set of relations:
\bea
\eta_2=\ex^{-{2i\pi\over3}}\eta_1,&&\qquad 
\eta_3=\ex^{{2i\pi\over3}}\eta_1,
\nonumber\\
\xi_2=\ex^{{2i\pi\over3}}\xi_1,&&\qquad
\xi_3=\ex^{-{2i\pi\over3}}\xi_1,\nonumber\\
 \zeta_2=\ex^{{-2i\pi\over3}}\zeta_1, &&\qquad
 \zeta_3=\ex^{{-4i\pi\over3}}\zeta_1.
\label{dcuatro}
\eea The relations among the $C_i$ are obtained by working out the  
contribution from the measure due to gravitational anomalies. The anomaly
comes from the  fields 
$\psi$, $\chi$, $\eta$, $\mu$, $\nu$, and implies the relations:
\be C_2=\ex^{-{i\pi\sigma\over6}}C_1,\qquad 
C_3=\ex^{-{i\pi\sigma\over3}}C_1.\
\label{dseis}
\ee

Denoting $C=C_1$, $\eta=\eta_1$, $\zeta=\zeta_1$ and $\xi=\xi_1$, one  has
the final result  for manifolds with $b_2^+ >1$ \cite{last}:
\begin{eqnarray} & &\langle {\rm exp}
(\sum_{a}\alpha_{a}I(\Sigma_{a})+\lambda {\cal  O})
\rangle \nonumber \\ &=&C \Bigg( {\rm exp} ({\eta\over2} v^2  +\lambda \xi)
\sum_{x} n_x {\rm exp}(\zeta v \cdot x)  \nonumber \\ &+&{\rm e}^{-{\pi i
\over 6}\sigma} {\rm exp}\Big( -{\rm e}^{-{\pi i \over 3}} ({\eta\over2}
v^2 +\lambda
\xi)
\Big) \sum_{x} n_x {\rm exp}({\rm e}^{-2\pi i \over 3} \zeta v 
\cdot x) \nonumber \\ &+&{\rm e}^{-{\pi i \over 3}\sigma} {\rm exp}\Big(
-{\rm e}^{{\pi i \over 3}} ({\eta\over2} v^2 +\lambda \xi)
\Big) \sum_{x} n_x {\rm exp}({\rm e}^{-4\pi i \over 3} \zeta v 
\cdot x) \Bigg), 
\label{fernanda}
\end{eqnarray} where unknown constants appear as in the pure
Donaldson-Witten case.  The generating function (\ref{fernanda}) verifies
a generalized form of the simple type condition (\ref{cochentaocho}):
\be 
\left({\partial^3\over{\partial\lambda^3}}-\xi^3\right)
\langle {\rm exp} (\sum_{a}\alpha_{a}I(\Sigma_{a})+\lambda {\cal  O})
\rangle =0.
\label{docho}
\ee

Unfortunately, the left-hand side of (\ref{fernanda}) is not known. Thus
we can not fix the constants $\eta$, $\xi$, $\zeta$ and $C$ as we did in
the case of Donaldson-Witten theory. That equation has to be regarded as a
prediction for those quantities. The result (\ref{fernanda}) suggests
that, as stated in the introduction,  moduli problems in four-dimensional 
topology can be  classified  in {\it universality classes} associated to
the  effective low-energy description  of the underlying physical theory.
One  important question that should be addressed is how large is the set
of moduli spaces which  admit a description in terms of Seiberg-Witten
invariants. It is very likely  that in the search for this set new types
of invariants will be found leading  to new universality classes. The case
considered in this section is the simplest of its kind. Other situations
should certainly be addressed.

\vfill
\newpage

\section{Final remarks}

We will end these lectures making several remarks. First of all we should
mention that there is a rich structure associated to twisted $N=4$
supersymmetric gauge theories. These theories can be twisted in three
non-equivalent ways \cite{yamron,marcus,australia,lozalabas}. Ideas based on duality have
been applied \cite{vafa} to one of the resulting twisted theories proving
invariance under the full duality group. The other two twisted theories,
as well as many other twisted theories coming from scale invariant $N=2$
supersymmetric gauge theories, should be addressed from a similar
perspective. It is expected that duality is realized for all these
theories.

In the context of generalized Donaldson-Witten theory it is worth to notice that
one of our assumptions can be released. We have assumed  in our discussion
that the manifold $X$ is spin. Of course, this is  only a simplifying
assumption. The theory can be also  defined on  general four-manifolds
using 
${\rm Spin}^c$-structures \cite{park,oko,oscar,pt}. It has  been recently
shown in the context of non-abelian monopoles that, from the point of view
of twisted $N=2$  supersymmetry, the inclusion of
${\rm Spin}^c$-structures corresponds to an  extended  twisting procedure
associated to the gauging of the baryon  number \cite{baryon}. 
 
It is also important to point out that the moduli space of non-abelian
monopoles has a natural $U(1)$ action  which  acts as a rotation on the
monopole fields. The fixed points of this  action  are essentially the
moduli space of ASD instantons and the moduli space  of abelian
Seiberg-Witten monopoles. This has opened the way to a  mathematical 
proof of the equivalence of both theories using localization
techniques \cite{pt,oko}, and some promising and concrete results in this 
direction  have been recently obtained \cite{okodos}. From the point of
view  of the Mathai-Quillen formalism the $U(1)$ action makes the bundle  
${\cal E}$ an $U(1)$-equivariant bundle and one can obtain a general
expression for  the equivariant extension of the Thom form in this
formalism \cite{eq}. For the  non-abelian monopole theory, the TQFT 
associated  to this extension is precisely twisted $N=2$
supersymmetric Yang-Mills coupled to one massive  flavor. This
result is promising for two reasons. Firstly, it  demonstrates that it is
possible to add mass-like parameters to this  type of theories while still
retaining the topological character of the  theory. Secondly, it is
tempting to think that the physics of the massive  theory could shed some
light on the localization problem.

Another important observation is the one made by Taubes \cite{taubes}
pointing out a relation between Seiberg-Witten invariants and Gromov
invariants. As mentioned in sect. 3,  Gromov invariants are the
topological quantities which appear in two-dimensional topological sigma
models. Therefore, what is involved here is a relation between TQFTs in
four dimensions and TQFTs theories in two dimensions. It is very likely
that an explanation of this relation based on physics will come from
string theory. In fact, it is tempting to speculate that string theory
will provide a new point of view to understand the relations among the
invariants associated to different moduli spaces. There are several
results which point into this direction. It is known, for example, 
that Donaldson theory shows up in certain compactifications of the
heterotic string \cite{harvey}. Similarly, other topological quantum field
theories, as the ones associated to  twisted $N=4$ supersymmetric gauge
theory can also be understood from a string theory
perspective \cite{vafastring}.

\vspace{3pc}

{\large\bf Acknowledgements}

\vspace{1pc}

We would like to thank M. Mari\~no for a critical reading of the 
manuscript and for many fruitful discussions on many aspects of the
subjects covered in these lectures. One of us (J.M.F.L.) would like to
thank the organizers of the workshop ``Trends in Theoretical Physics", for
inviting him to deliver these lectures and for their warm hospitality. We
acknowledge funds provided by the European Commission,
which supports the collaboration network `CERN-Santiago de Compostela-La
Plata' under contract C11$^*$-CT93-0315, for making possible the
organization of the workshop. This work is supported in part by DGICYT
under grant PB93-0344.

\end{document}